\documentclass[12pt,preprint]{aastex}
\received{}
\date
\accepted{}
\slugcomment{}

\begin{document}

\title{AGN Black Hole Masses and Bolometric Luminosities}

\author{Jong-Hak Woo}
\affil{Department of Astronomy, Yale University, P.O. Box 208101, New Haven, CT 06520-8101}
\email{jhwoo@astro.yale.edu}
\author{C. Megan Urry}
\affil{Department of Physics and Center for Astronomy and Astrophysics, Yale University, P.O. Box 208121, New Haven, CT 06520-8121}
\email{meg.urry@yale.edu}

\begin{abstract}
Black hole mass, along with mass accretion rate, is a 
fundamental property of active galactic nuclei. Black hole mass 
sets an approximate upper limit to AGN energetics via the
Eddington limit.
We collect and compare all AGN black hole mass estimates 
from the literature; these 177 masses are mostly based on the virial
assumption for the broad emission lines, with the broad-line region
size determined from either reverberation mapping or 
optical luminosity. 
We introduce 200 additional black hole mass estimates based on properties
of the host galaxy bulges, using either the observed stellar velocity
dispersion or using the fundamental plane relation to infer $\sigma$;
these methods assume that AGN hosts are normal galaxies.
We compare 36 cases for which black hole mass has been generated 
by different methods and find, for individual
objects, a scatter as high as a couple of orders of magnitude.
The less direct the method, the larger the discrepancy with other
estimates, probably due to the large scatter in the underlying 
correlations assumed.
Using published fluxes, we calculate bolometric luminosities for 234 AGNs
and investigate the relation between black hole mass and luminosity.
In contrast to other studies, we find no significant correlation
of black hole mass with luminosity, other than those induced by
circular reasoning in the estimation of black hole mass.
The Eddington limit defines an approximate upper envelope to the
distribution of luminosities, but the lower envelope depends entirely
on the sample of AGN included. 
For any given black hole mass, there is a range in Eddington ratio 
of up to three orders of magnitude. 
\end{abstract}

\section{Introduction}

Black holes have been the leading candidate to power the central engines
in AGN for over three decades (Lynden-Bell 1969), but direct
evidence for their presence has been elusive. 
In nearby galaxies, spatially resolved kinematics have provided strong
evidence for the ubiquity of nuclear black holes, with dynamical
black hole detections reported for 37 galaxies (Kormendy \& Gebhardt 2001).
Such observations are available only for a handful of the nearest AGN
(Harms et al. 1994, Miyoshi et al. 1995, Greenhill et al. 1996). 

Black hole mass, along with mass accretion rate, 
is a fundamental property of AGN.
Via the Eddington limit, 
a maximum luminosity for the idealized case of spherical accretion
($L_{Edd} = 1.25 \times 10^{38} \times M_{BH}/M_{\odot}$ ergs s$^{-1}$), 
the black hole mass sets an approximate upper limit to AGN energetics. 
It is also the integral of the accretion history of the AGN.
However, direct kinematic observations of the black hole mass are limited by
finite spatial resolution (a typical AGN at redshift 2 would 
require nano-arcsecond resolution to probe the sphere of influence 
of the black hole), not to mention that scattered light from the 
bright central source dilutes any kinematic signal from orbiting material.

For these reasons, various less direct methods for estimating black hole mass 
have been devised.
One set of methods (\S\S~2.1,2.2) assumes the broad-line region (BLR) 
is gravitationally bound by the central black hole potential, so that
the black hole mass can be estimated from the orbital radius
and the Doppler velocity.
The reverberation mapping technique utilizes the time lag between
continum and emission lines to derive the distance of the BLR from
the black hole (Blandford \& McKee 1982, Peterson 1993).
About three dozen AGN black hole masses have been measured using
this technique.
A less costly alternative is to infer the BLR size from the
optical or ultraviolet luminosity (McLure \& Dunlop 2001, Vestergaard 2002),
with which it is correlated, at least over
a limited range of luminosities (Kaspi et al. 2000).

A different approach to estimating black hole mass is to exploit
the correlation, seen in nearby normal galaxies, 
between black hole mass and stellar velocity dispersion, $\sigma$ 
(Ferrarese \& Merritt 2000, Gebhardt et al. 2000a).
If AGN host galaxies are similar to non-active galaxies, 
this correlation should hold also for them.
Since stellar velocity dispersion measurements are still difficult
for higher redshift AGN, the stellar velocity dispersion 
can possibly be inferred from effective radius and central surface brightness 
assuming AGN host galaxies occupy the same fundamental plane 
as ordinary ellipticals (O'Dowd et al. 2002). 

Some previous studies have found a tight relation between 
mass and luminosity in AGN
(Dibai 1981, Wandel \& Yahil 1985, 
Padovani \& Rafanelli 1988, Koratkar \& Gaskell 1991, Kaspi et al. 2000);
however, the scatter is large when the black hole masses
are restricted to the most reliable estimates
(from reverberation mapping).
One might have expected a correlation between
AGN black hole mass and luminosity since the
Eddington luminosity is proportional to black hole mass,
but if there is a range in accretion rates and/or efficiencies,
the relation will be weaker. 

In this paper, we collect and compare all AGN black hole mass estimates 
from the literature, and we make new black hole mass estimates from
stellar velocity dispersions (\S~2). 
We calculate bolometric luminosities for these same AGN to
investigate their mass--luminosity relation, and look for trends
of Eddington ratio with luminosity (\S~3).
Table~\ref{T_SUM} summarizes the number of black hole mass estimates 
from the various methods.
We use $H_0=75$~km s$^{-1}$ and $q_0=0.5$ throughout this paper.

\section{Black Hole Masses in AGN}

Very few black hole masses in AGN have been measured with spatially
resolved dynamics near the central black hole which is the preferred
method for estimating black hole mass in nearby (inactive) galaxies.
The two cases in which this has been done with maser kinematics
(NGC~1068 and NGC~4258) are listed in Table 2. Remaining black hole
masses are determined with less direct methods.

\subsection{Masses from the Virialized Motion}

Assuming that broad-line clouds are virialized,
for which there has been increasing evidence
(Krolik et al. 1991, Wandel et al. 1999, cf. Krolik 2001), 
the black hole mass can be estimated:
\begin{equation}
M_{BH} = R_{BLR}~ v^{2}~G^{-1} ~. 
\label{virial}
\end{equation}

The virial assumption may not be correct, however;
radiation pressure and/or magnetic fields may contribute
significantly to the dynamics (Krolik 2001), and 
outflows or winds could cause the observed line widths 
to exceed those induced by the black hole potential alone. 
In these cases the black hole mass calculated from
Eq.~\ref{virial}
would be overestimated.

\subsubsection{Reverberation Mapping Estimates}

In reverberation mapping,
the BLR size is estimated from the time lag between
the ionizing continuum and the broad-line strength
(Peterson 1993).
To date, 36 AGN black hole masses have been measured from 
combining reverberation-mapped BLR sizes with
broad-line velocities (Wandel et al. 1999, Ho 1999, 
Kaspi et al. 2000, Onken \& Peterson 2002). 
These are listed in Table~\ref{T_rev}, along with
the redshifts, bolometric luminosities, and published
AGN types. 

\begin{figure}
\plotone{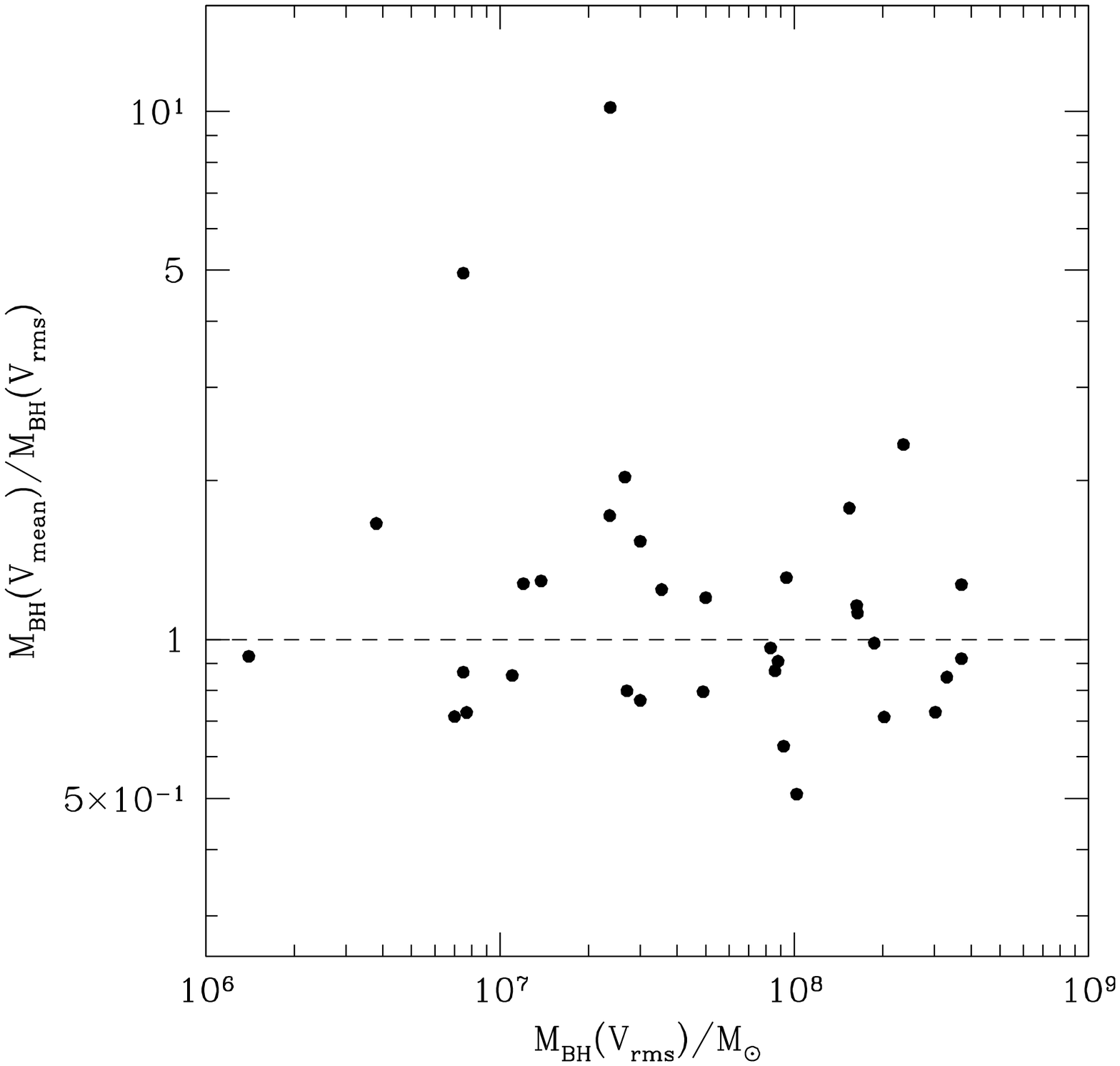}
\caption{Comparison of black hole masses calculated for different
FWHM estimates --- mean FWHM and FWHM of the rms spectrum ---
for the 34 reverberation-mapped AGN of Kaspi et al. (2000).
The difference in black hole mass for the same AGN is as 
large as an order of magnitude.
\label{REV}}
\end{figure}

Contributing to the uncertainty in the black hole
mass estimation are the BLR orbits and velocities assumed.
The broad line velocity can be determined from
the observed spectra, either as the mean of the 
FWHM derived from each line or as the FWHM from the
root mean square (rms) spectrum (Peterson et al. 1988).
Kaspi et al. (2000) showed that the two velocity estimates are
similar; however, the difference between the two gives black hole 
mass uncertainties as large as a factor of ten (Figure~\ref{REV}).

Assumptions about the orbital shape and inclination
of the broad-line clouds introduce additional uncertainties. 
An isotropic distribution with random inclinations is often 
assumed for the broad-line clouds, in which case velocity is derived 
from Equation~(2) with $f=\sqrt{3}/2$ 
(Netzer 1990): 

\begin{equation}
v = f \times FWHM ~.
\label{width}
\end{equation}

However, the random orbits assumption may not be valid 
for quasars. McLure \& Dunlop (2001) reproduced the FHWM 
distribution of Seyferts and quasars with two disc components, 
and determined that the average relationship between observed 
FWHM and actual orbital velocity corresponds to $f= 3/2$.
Thus for the same AGN,
the black hole mass estimates in McLure \& Dunlop (2001) 
are factor of 3 larger than those of Kaspi et al. (2000).
Considering orbital shape alone, the full range of uncertainty 
in mass appears to be 2 orders of magnitude, 
from $f=3/2$ to $\sim$ 200 (Krolik 2001).

In Figure~1 we compare 34 reverberation-mapped black hole masses 
calculated for two different estimates of the broad-line velocities 
(Kaspi et al. 2000). 
The derived black hole masses for a given object
differ by less than an order of magnitude, making 
reverberation mapping one of the more
robust techniques for estimating AGN black hole masses.
It is however resource intensive, 
time consuming, and not applicable to most AGN (those
without broad lines). 
Consequently, relatively few AGN black hole 
masses have been well estimated.

\subsubsection{Black Hole Mass Estimates using the BLR Size -- 
Luminosity Relation}

Since reverberation mapping is a laborious process,
alternative ways of deriving the BLR size are of interest.
Several authors have noted that $R_{BLR}$ (where known from 
reverberation mapping) appears to correlate with UV/optical luminosity
(Koratkar \& Gaskell 1991, Kaspi et al. 1996, Wandel et al. 1999, 
Kaspi et al. 2000). 
The proportionality has been reported as $L_{opt}^{1/2}$ (Wandel et al. 1999),
which corresponds to a constant ionization parameter,
but in the most recent studies appears to be
$R_{BLR}\propto L_{5100{\rm \AA}}^{0.7}$ (Kaspi et al. 2000, Vestergaard et al. 2002;
cf. McLure \& Jarvis 2002).
Using this relation and assuming random isotropic orbits
($f= \sqrt{3}/2 $ in Eq.~\ref{width}), we obtain:
\begin{equation}
M_{BH} = 4.817 \times (\frac{\lambda L_{\lambda}(5100{\rm \AA})}{10^{44}~{\rm ergs s^{-1}}})^{0.7} \times
~(FHWM)^{2} ~.
\label{Lopt}
\end{equation}

\begin{figure}
\plotone{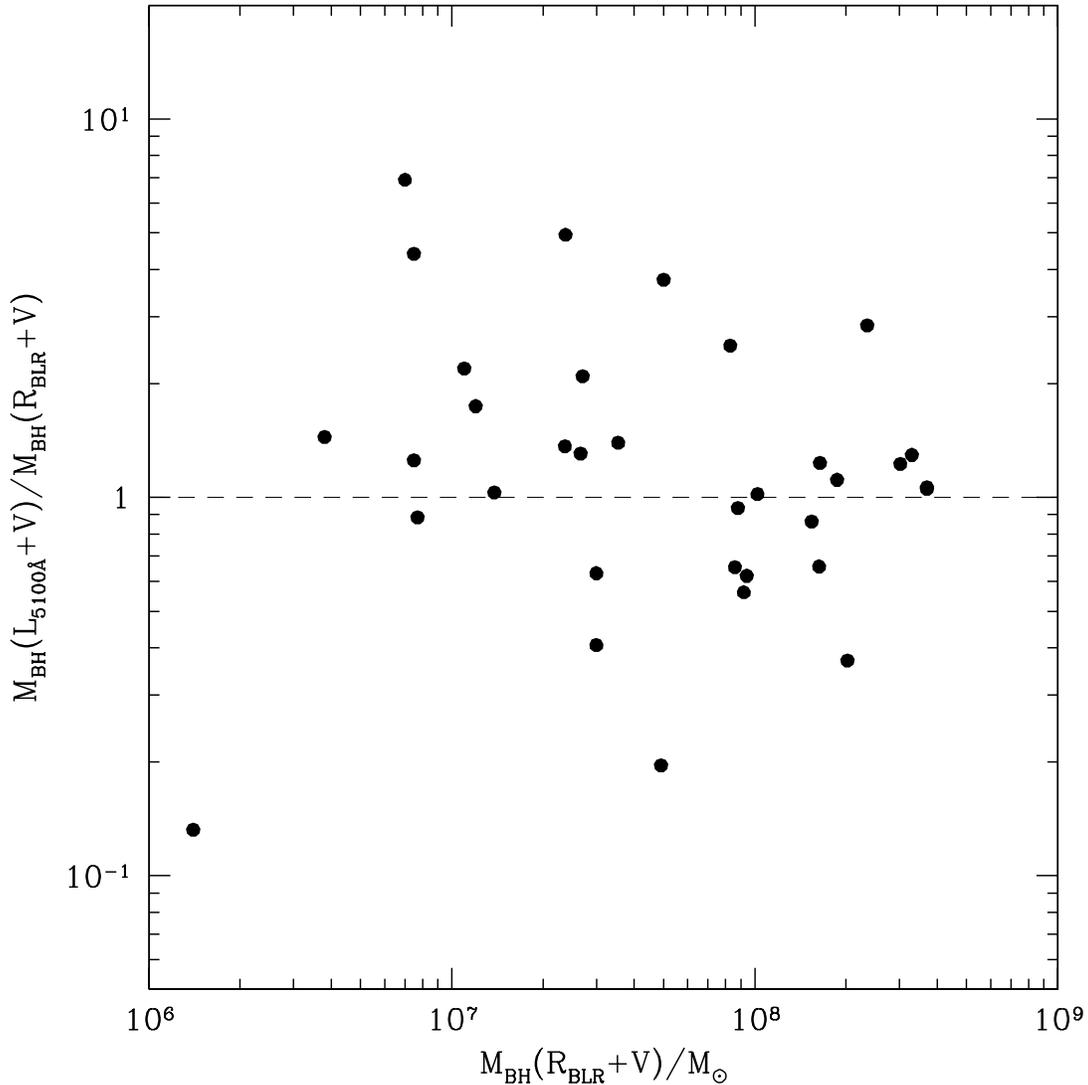}
\caption{Comparison of black hole masses calculated using 
two different estimates of broad-line region (BLR) size --- 
from reverberation mapping and from the
$R_{BLR}$ -- $L_{5100{\rm \AA}}$ relation of Kaspi et al. (2000) ---
combined with the rms velocity of the $H\beta$ line
(assuming $f= \sqrt{3}/2 $ in Eq.~\ref{width}, corresponding
to random isotropic orbits).
Relative uncertainties are as large as an order of magnitude,
and come mainly from the large scatter in the size--luminosity
relation.
The unknown orbits add another factor of 3 or more
uncertainty in the black hole mass (not represented in this plot).
\label{OPT_REV}}
\end{figure}

There is large scatter in the 
$R_{BLR}$ -- $L_{5100 {\rm \AA}}$ correlation
(e.g., Figure~7 of Kaspi et al. 2000), 
and it has been established
only over a limited range of luminosities, hence it yields
correspondingly uncertain black hole masses.
We list these values in Table~\ref{T_opt}, 
along with the redshift, 
bolometric luminosity, and AGN type,
and in Figure~\ref{OPT_REV} we compare them
to all available reverberation mapping estimates.
The differences range up to an order of magnitude,
with an rms difference of 0.50 in the log of the ratio.

If optical luminosity is well correlated with bolometric luminosity,
the fitted correlation of Kaspi et al. (2000) leads to a
precise relation between black hole mass and bolometric luminosity
(something we would like to investigate rather than assume). 
The Eddington ratio (i.e., the ratio of bolometric 
luminosity to Eddington luminosity) would then depend on
bolometric luminosity to the 0.3 power.

Although there are some concerns, black hole mass estimates 
with this method remain important given the difficulty of 
more accurate estimates and the relatively small number of AGN
for which any black hole mass estimates have been made.
Thus, we collected all such black hole mass estimates 
available in the literature (26 from McLure \& Dunlop
2001, 3 from Laor 2001, 80 from Gu et al. 2001, 30 from Oshlack, Webster \& Whiting 2002), 
re-computed using Eq.~\ref{Lopt}
for consistency with our cosmology.

\subsection{Black Hole Mass from Stellar Velocity Dispersion}

In nearby galaxies there is apparently a close connection
between the central black hole and the bulge kinematics. 
Specifically, black hole mass (determined from spatially 
resolved kinematics) correlates well with stellar velocity 
dispersion, 
as $M_{BH}\propto \sigma^{3.75}$ (Gebhardt et al. 2000a) 
or $M_{BH}\propto \sigma^{4.8}$ (Ferrarese \& Merritt 2000).
From the collective analysis by Tremaine et al. (2002): 
\begin{equation}
M_{BH} = 1.349 \times 10^{8} M_{\odot} (\sigma/200 {\rm km~s}^{-1})^{4.02} ~.
\label{sig}
\end{equation}

AGN host galaxies appear to be very much like normal galaxies.
This is particularly well established for radio-loud AGN,
whose host galaxies follow the usual Kormendy relation
(Taylor et al. 1996; McLure et al. 1999; Urry et al. 2000;
Bettoni et al. 2001). 
Present data on host galaxies are in accord with the 
``grand unification'' hypothesis, 
suggested on other grounds, that
AGN are simply a transient phase of normal galaxies 
(Cavaliere \& Padovani 1989).
Therefore it is reasonable to expect that the 
same $M_{BH}$--$\sigma$ correlation should be present 
in AGN host galaxies, in which case 
we can use Eq.~\ref{sig} to infer black hole mass.
Gebhardt et al. (2000b) and Ferrarese et al. (2001) 
estimated black hole masses in this way for a few Seyfert galaxies
(7 and 6 respectively), and found good agreement with
reverberation mapping values.

\subsubsection{From Direct Measurement of Stellar Velocity Dispersion}

An increasing number of AGN have published measurements
of stellar velocity dispersion. 
Black hole masses calculated from $\sigma$ have been published for
21 Seyferts (Wu \& Han 2001) and
12 BL Lac objects (Falomo et al. 2002; Barth, Ho \& Sargent 2002);
we rescaled these to our cosmology as needed.
From the literature we collected velocity dispersions 
for an additional 108 AGN (36 Seyfert galaxies and 72 radio galaxies), 
and calculated their black hole masses according to Eq.~\ref{sig}.
All 141 black hole masses are presented in Table~\ref{T_sig}.

\begin{figure}
\plotone{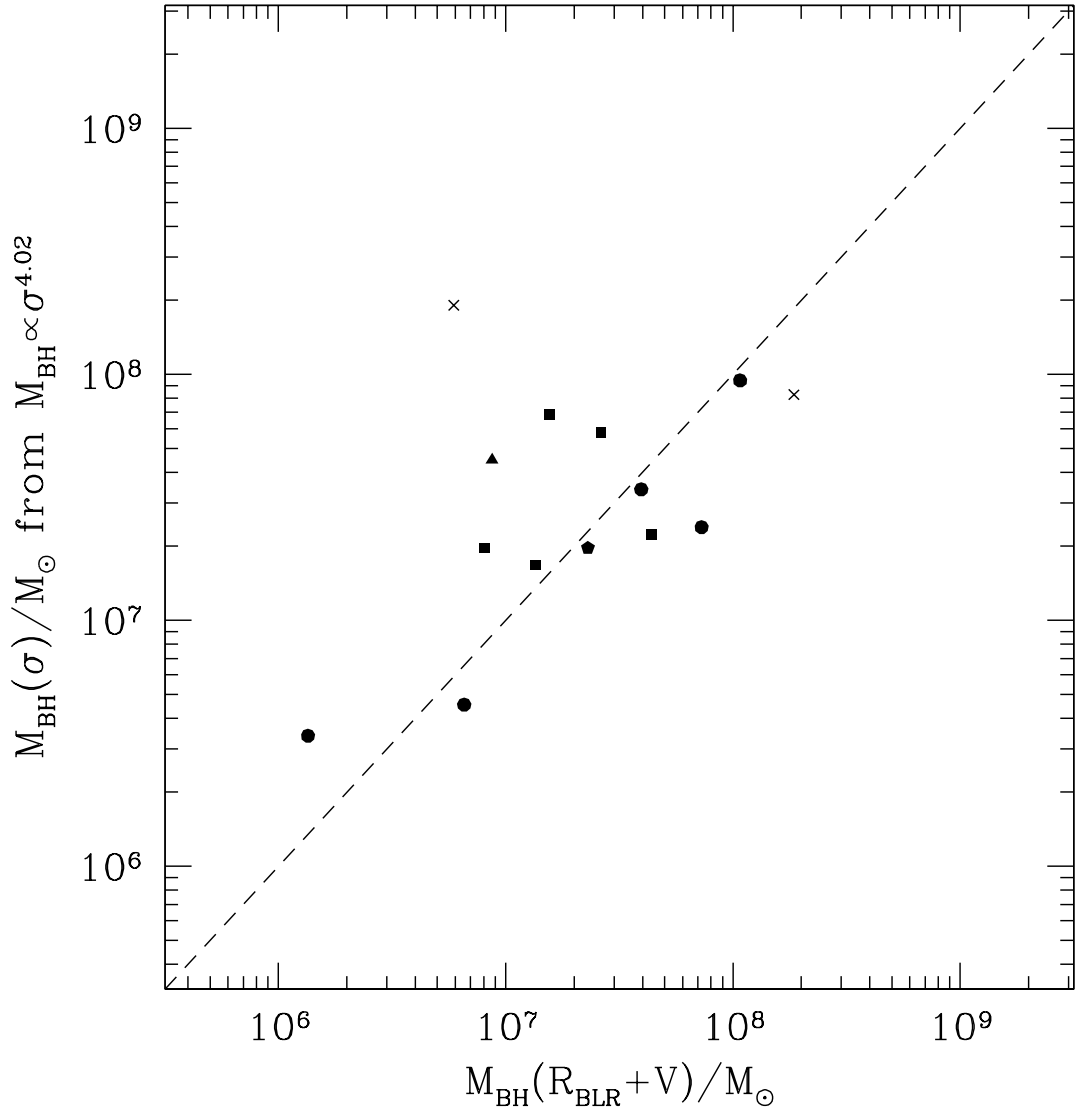}
\caption{
Comparison of two completely independent estimates of
black hole mass, one from the stellar velocity dispersion correlation,
$M_{BH} \propto \sigma^{4.02}$,
and the other from reverberation mapping.
Apart from one discordant object, IC~4329, the two masses agree well,
with dispersion less than 50\%.
Reverberation masses are based on the values in 
Kaspi et al. (2000; log mean of two values from 
rms and mean velocity),
Ho (1999; {\it triangle}),
and Onken \& Peterson (2002; {\it crosses}).
Stellar velocity dispersion masses are from Nelson (1995; {\it squares}),
Ferrarese et al. (2001; {\it circles}), Oliva et al. (1995; {\it triangle}),
Di Nella et al. (1995; {\it pentagon}), and Oliva et al. (1999; {\it crosses}).
\label{SIG_REV}}
\end{figure}

For 14 Seyfert galaxies both velocity dispersions and
reverberation-mapped BLR sizes are available. 
In Figure~\ref{SIG_REV} we compare the two associated
black hole mass estimates. 
They agree relatively well, with scatter much
less than an order of magnitude. 

\subsubsection{From Indirect Estimates of Stellar Velocity Dispersion}

Stellar velocity dispersions are not extensively
known for AGN host galaxies, nor are they easy to measure, 
particularly at higher redshift. 
However, by the same ``grand unification'' of host galaxies
with normal galaxies, we can infer the velocity dispersions
(albeit with additional scatter) from the morphological
parameters of the bulge: $r_e$, the effective radius, and $\mu_e$, the
surface brightness at that radius.
These have been very well measured for more than 100
AGN using the excellent
spatial resolution of the Hubble Space Telescope\footnote{Based
on observations made with the NASA/ESA Hubble Space Telescope, 
obtained at the Space Telescope Science Institute,
which is operated by the Association of Universities 
for Research in Astronomy, Inc., under NASA contract 
NAS 5-26555. These observations are associated with proposals
\# 5849, 5938, 5939, 5949, 5957, 5974, 5982, 5988, 6303, 6361, 6363, 6490, 6776, 7893.}
(HST), which yields more robust results than
observations in typical ground-based seeing.

Thus, at least for radio-loud AGN, black hole mass can be
derived from $r_{e}$ and $\mu_{e}$ (O'Dowd et al. 2002).
If sufficiently accurate, this would be an extremely valuable 
method since the required imaging data are much easier to obtain 
than $\sigma$, and such a method could be applied widely 
and at higher redshift than the direct method.

Using this method, we estimate 59 new black hole masses 
for 45 BL Lac objects, 10 radio galaxies,
and 4 radio-quiet AGN, all of which have host galaxies
detected with HST.
Surface brightnesses and effective radii from
Urry et al. (2000) and Dunlop et al. (2002) are used to derive 
stellar velocity dispersion via fundamental plane relation of 
Jorgensen et al. (1996):

\begin{equation}
\log r_{e} = 1.24~ log \sigma -0.82~log <I_{e}> + 0.2132~z - 0.00131-C ~.
\label{fp}
\end{equation}

\noindent
Here, $C=0.176$ for cosmological correction to $H_0$=75 km s$^{-1}$. 
Black hole masses are then estimated using 
Eq.~\ref{sig}.
Morphological parameters and derived black hole masses 
are given in Table~\ref{T_fp}. Bolometric luminosity is
not straightforward to derive for most of these objects
because of beaming and obscuration.

\begin{figure}
\plotone{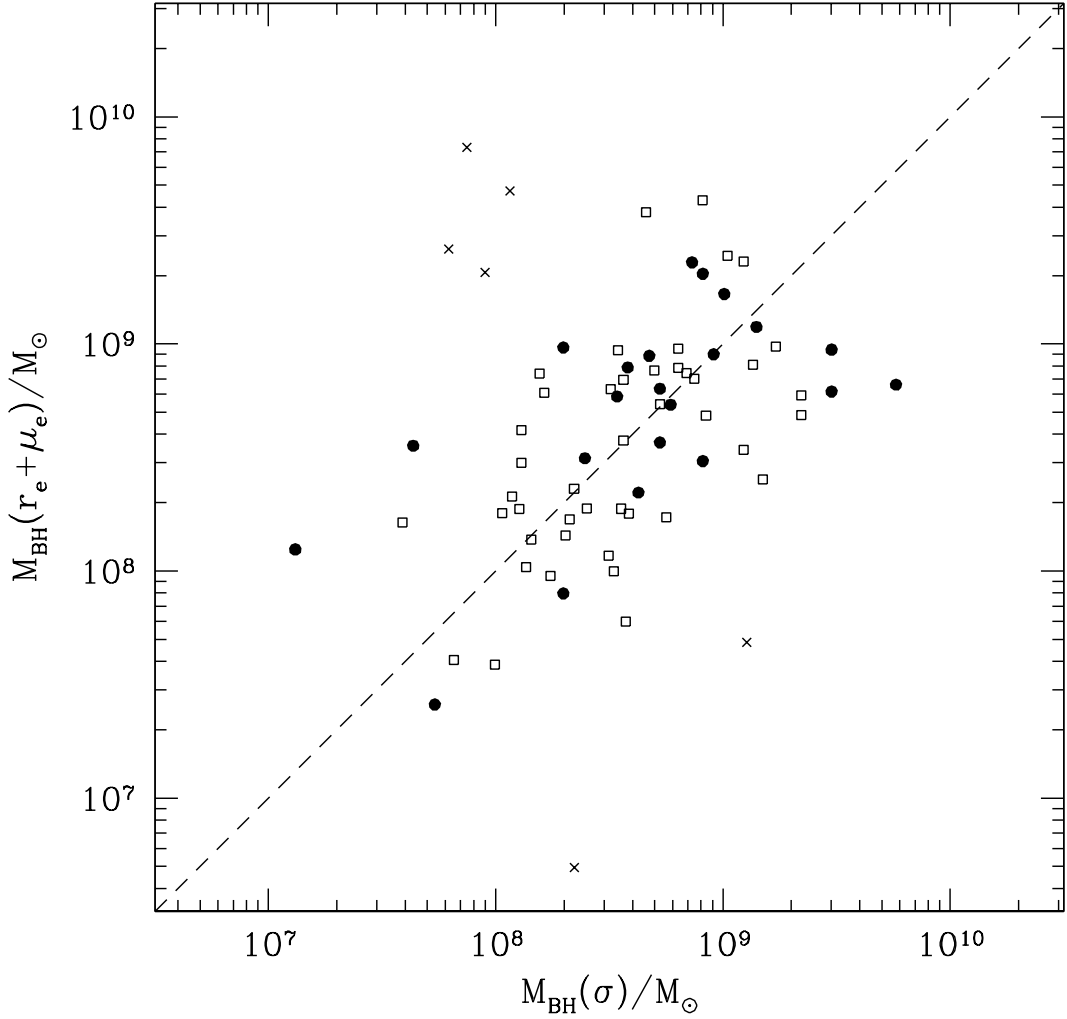}
\caption{
Black hole masses estimated from 
the correlation with stellar velocity dispersion,
for the Bettoni et al. (2001) sample of radio galaxies.
The plot compares $M_{BH}(r_e + \mu_e)$, 
derived from an indirect estimate 
of $\sigma$ based on measured $r_{e}$ and $\mu_{e}$)
and the fundamental plane relation,
to $M_{BH}(\sigma)$, derived from 
direct measurements of the stellar velocity dispersion.
Measurements of $\sigma$ include a homogeneous set 
of 22 new measurements presented by Bettoni et al. ({\it filled circles}) 
and another 50 measurements ({\it open squares}) assembled
by Bettoni et al. from the literature; the latter have larger
scatter probably because they had to be transformed 
in color (from V to R) and corrected for different apertures.
Apart from 6 outliers ({\it crosses}),
most values agree well, with an rms dispersion of less than a factor of 4.
\label{FP_SIG}}
\end{figure}

To test the accuracy of this fundamental plane method for
estimating black hole mass, we considered 72 radio 
galaxies for which all three parameters of the fundamental plane
are measured (Bettoni et al. 2001).\footnote{Table~3 of Bettoni et al. (2001)
apparently lists $r_{e}$ values in arcsec rather than kpc (Barth et al. 2002).}
Figure~\ref{FP_SIG} shows the comparison of
black hole masses derived indirectly from $\mu_e$ and $r_e$
with those derived directly from $\sigma$.
(This is in effect an unusual projection of the fundamental plane.)
Points are coded to highlight the homogeneous data of
Bettoni et al. (filled circles), which are more tightly
correlated than the additional heterogeneous data (open squares and crosses)
collected by them.
The six most extreme outliers are marked with crosses.
The mean black hole masses determined by the two methods
agree to within 10\%, while the rms scatter is
a factor of 4 or so (slightly higher for the heterogeneous 
data than for the homogeneous data).

Although the fundamental plane method introduces additional
scatter compared to direct measurement of stellar velocity dispersion,
estimating black hole masses in this way is so far
one of the few ways to infer AGN black hole mass for high redshift AGN
(perhaps the only method for AGN that lack broad emission lines).
Of course, the underlying assumption of 
``grand unification'' of AGN and galaxies
remains untested, particularly at high redshift.

\section{Bolometric Luminosity and Black Hole Mass}

\subsection{Bolometric Luminosity of AGN}

Bolometric luminosity of AGN is sometimes approximated 
from optical luminosity, since integration of the 
spectral energy distribution (SED), which spans
many decades in wavelength, is usually hampered by
lack of wavelength coverage and by variability. 
Here we are able in many cases to determine bolometric luminosity
by integrating all available flux points in the SED.
This is particularly important given the role of 
optical luminosity in deriving some black hole masses,
otherwise correlations between $M_{BH}$ and $L_{bol}$ 
can be induced.

For 234 of the 377 AGN for which black hole mass has 
been estimated in the Tables, we were able to determine
bolometric luminosity. 
The other 143 objects are radio galaxies and BL Lac objects, 
for which obscuration and beaming are significant.
For 82 of the 234, there are numerous published fluxes
from ultraviolet to far-infrared wavelengths, which we
collected using the NED database.\footnote{
The NASA/IPAC Extragalactic Database (NED) is operated by the Jet 
Propulsion Laboratory, California Institute of Technology,
under contract with the National Aeronautics and Space Administration.}
Multiple observations for the same band were simply averaged, 
and the Galactic extinction law (Cardelli, Clayton \& Mathis 1989) was used 
to correct for dust (with $A_{V}$ also taken from NED). 
We then integrated these SEDs directly
to get the bolometric luminosity.

For the remaining AGN,
mostly quasars at relatively high redshift, 
sufficient flux points were unavailable. 
In 152 cases, including most of the luminous quasars,
we obtained the bolometric luminosity by
fitting the average SED for that AGN type to 
the available flux points.
Average SEDs are from various sources: radio-loud and radio-quiet quasar SEDs are from 
Elvis et al. (1994); Seyfert 1 SEDs are from Mas-Hesse et al. (1994);
and Seyfert 2 SEDs are from Schmitt et al. (1997).
Optical flux was corrected for Galactic extinction using 
individual reddening values from NED.
We note that the bolometric luminosities are roughly
10 times the optical luminosity (precisely, in the case of
SED fitting for quasars, and within a factor of 5-6 in the
case of direct integration of the SEDs).

Bolometric luminosities for a total of 234 AGN 
are given in Tables~\ref{T_kin}, \ref{T_rev}, \ref{T_opt} and \ref{T_sig}. 
The associated black hole masses were estimated 
as follows: 
2 from maser kinematics,
36 from broad-line widths plus reverberation mapping, 
139 from broad-line widths plus the $L_{5100{\rm \AA}}$ - $R_{BLR}$ relation, 
and 57 from the $M_{BH}$ - $\sigma$ relation.

\begin{figure}
\plotone{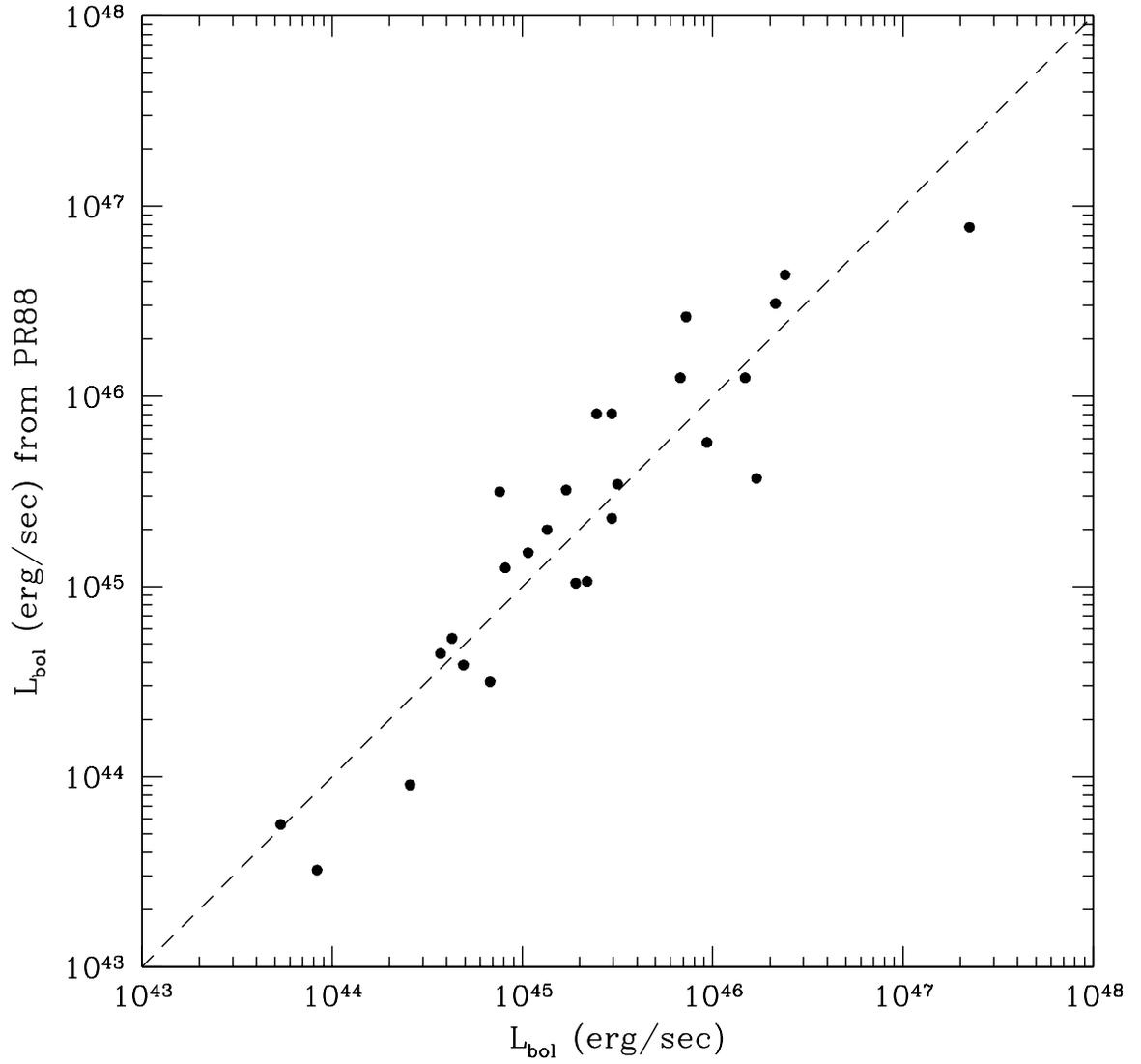}
\caption{Comparison of bolometric luminosity measurements
from present paper to those of Padovani \& Rafanelli 1988, 
for the 26 AGN found in both samples.
The two values are consistent;
the very slightly smaller values found by Padovani \& Rafanelli 
are due to the more limited spectral range over which they
integrated the flux.
\label{LBOL}}
\end{figure}

In order to check our bolometric luminosity measurements,
we compare them with previous estimates by 
Padovani \& Rafanelli (1988), 
who integrated available optical to far-infrared fluxes for
58 Seyfert galaxies and quasars.
Twenty-six AGN in the Padovani \& Rafanelli sample
have bolometric luminosities estimated here;
we rescaled the former values
to $H_0=75$~km s$^{-1}$ simply by multiplying by 4/9 
($H_0=50$~km s$^{-1}$, $q_0=0$ in their calculation). 
The comparison is shown in Figure~\ref{LBOL}.
The two estimations agree well although the
Padovani \& Rafanelli values may be systematically lower
due to the more limited spectral range in their calculation.

\subsection{The Black Hole Mass -- Luminosity Relation }

We now compare bolometric luminosity with black hole mass.
Figure 6a includes only the 36 reverberation-mapped
quasars and Seyfert galaxies, and Figure 6b includes the
57 Seyfert galaxies for which black hole mass was
estimated from observed stellar velocity dispersion.
There is large scatter and little correlation between
bolometric luminosity and black hole mass.
For a given black hole mass, the bolometric luminosity ranges
over more than two orders of magnitude.
Figure 6c shows the mass--luminosity plot for AGN
with black hole masses that were derived from
optical luminosity and broad-line velocity
(McLure \& Dunlop 2001, Laor 2001, Gu et al. 2001,
Oshlack et al. 2002).
Even here there not much more of a correlation,
although one will appear if optical and bolometric
luminosities are well correlated.
That is, since black hole masses for these AGN were
derived from $L_{5100{\rm \AA}}$,
the slope indicated by the solid line is implied if
$L_{bol}$ is proportional to $L_{5100{\rm \AA}}$.

Figure~\ref{ML_ALL} shows the mass--luminosity relation
for all 234 AGN. Even more clearly than in Figure~6,
there is hardly any trend of luminosity with black hole mass.
For a given AGN black hole mass, the
bolometric luminosity ranges over at least two,
and as much as four, orders of magnitude.
The Eddington ratio must span a similarly large range.
The Eddington ratio does define an approximate (but not
hard) upper limit to the distribution of luminosities;
that is, points are missing from the upper left region above
the dotted line, in fact previously noted 
by McLeod, Rieke \& Storrie-Lombardi (1999).
The lack of points in the lower right, however, is a
selection effect: this part of the diagram gets filled
in simply by including lower luminosity AGN, continuously
down to galaxies.
Among the low-luminosity objects with large black
holes are the radio galaxies and BL Lac objects
for which we do not have good estimates of bolometric
luminosity (cf. O'Dowd et al. 2002);
the box indicates the approximate region they occupy,
calculated from the observed luminosities of BL Lacs
using the family of SEDs from Fossati et al. (1998)
and correcting for beaming factors in the range 3-10
(Dondi \& Ghisellini 1995).

AGN lore has it that the Eddington ratio is
0.1-1 for high-luminosity sources and an order of magnitude 
or more smaller for low-luminosity sources.
Our sample of AGN spans 5 decades in bolometric luminosity so
we should be very sensitive to any such trends. 
In Figure~\ref{L_ER} we plot Eddington ratio versus bolometric 
luminosity (top panel).
At most luminosities, the Eddington ratio spans two
decades or so, except at the very highest luminosity.
There appears to be a deficit of high luminosity objects
with low Eddington ratios (i.e., with black holes
in the range $10^{8} < M_{\rm BH} / M_\odot < 10^{10}$).
However, these include some of the radio
sources for which we do not have good bolometric luminosities
(see Table~\ref{T_fp}). Furthermore, if more massive black
holes are rare (i.e., there is a steep mass function),
they would on average be found at high redshift, yet
low-luminosity radio sources at high redshift are
excluded from flux-limited samples.
There is also a deficiency of points in the upper left
corner of the plot; these would be AGN with luminosities
of $\lesssim 10^{44}$~ergs s$^{-1}$ and black hole masses less
than $10^6$~$M_{\odot}$. (Note that low-luminosity AGN may be more difficult
to detect because of dilution by host galaxy light.)
Thus there is no immediate evidence of any
real trend in Eddington ratio with luminosity.

We also plot Eddington ratio versus black hole mass
(bottom panel). 
Again, there are no clear trends that cannot be
explained by sample selection effects. For example,
objects with luminosities below $10^{43}$~ergs s$^{-1}$ are
not called Seyfert galaxies or quasars and thus do
not appear in this diagram. (One could add them, and
they would fill in the lower left corner of the plot.)
AGN with luminosities greater than $10^{47}$~ergs s$^{-1}$ are
rare and thus probably too distant, on average, to have
black hole mass estimates.
With such a heterogeneous sample, we hesitate to make
any strong statements, but certainly we see only very
weak trends or correlations, and those are quite plausibly
induced by sample selection effects.

We can see this by plotting
the Eddington ratio versus redshift (Figure~\ref{Z_ER}).
Again there is little if any trend. 
High Eddington ratio objects ($L_{bol}/L_{Edd} \gtrsim 1$)
are perhaps missing at low redshift, but this can be
explained as a volume effect (i.e., given the steep luminosity
function of AGN, one has to survey a large volume
to find a relatively rare high-luminosity AGN).
More obviously, low Eddington ratio objects
$L_{bol}/L_{Edd} \gtrsim 1$) are absent at high
redshift, and this is partly a flux limit issue,
since low-luminosity AGN fall out of samples at
high redshift. Thus any trends that do appear to the
eye in this plot are explained by obvious selection
effects.

\subsection{Black Hole Mass and Radio Luminosity}

Finally, we look at radio luminosity versus black hole mass
(Fig.~\ref{L5GHZ}, top panel)
since previous reports have suggested there is a correlation
between the two (McLure et al. 1999; Lacy et al. 2001),
although more recent investigations have not found such a correlation
(Ho 2002; Oshlack et al 2002).
Again, there is little evidence of a correlation, particularly
given the missing low-luminosity sources like BL Lacs that
do appear to have high black hole masses (and thus should
help fill in the lower right corner of the plot).
Very low-luminosity AGN 
($L < 10^{23}$~W/Hz) with massive black holes may be missing,
though this is hard to quantify given the missing BL Lacs
and radio galaxies. 

To further investigate this point, we consider radio-loudness.
There have been suggestions that black hole mass is a factor
in radio loudness, such that $R>1$ 
($R \equiv L_{5~{\rm GHz}} / L_{5000~{\rm \AA}}$)
requires $M_{BH} \gtrsim 10^9~M_\odot$ (Laor 2000).
In Figure~\ref{L5GHZ} we plot radio loudness versus black hole mass 
for the same objects (bottom panel).
The radio-loud AGN have a very broad distribution of masses,
so there clearly is no threshold effect.
In the radio-quiet regime ($R < 1$), the distribution of masses
is narrower, with no black holes masses greater than 
$M_{BH} \gtrsim 10^9~M_\odot$.
We note that almost all of the high-mass black holes
are estimated from the optical luminosity method;
that these occur in radio-loud AGN, therefore, could
be explained if an appreciable fraction of the optical
luminosity is beamed. 
If instead the absence of high-mass radio-quiet AGN is real,
this would be a very significant 
distinction between the radio-quiet and radio-loud AGN.
However, given the heterogeneous sample discussed here,
the absence of evidence of these objects is not evidence
of their absence, and more work will be required on this point.

\section{Summary and Conclusions}

We estimated and/or collected from the literature
black hole masses for 377 AGN, obtained with various methods.
These span a range of nearly 4 orders of magnitude,
from $10^6~M_\odot$ to $7 \times 10^{9}~M_\odot$.
Direct comparisons suggest that 
reverberation mapping and stellar velocity dispersion 
give reliable black hole mass estimates --- within factors
of a few --- while using optical luminosity to infer 
broad-line size or using the fundamental plane to 
infer velocity dispersion leads to somewhat larger uncertainties. 
In the case of virial estimates (reverberation mapping,
optical luminosity, or other), additional uncertainties
enter through the unknown orbits and the possible non-virial 
motions of the line-emitting gas.

We estimated bolometric luminosities for most of the
AGN, apart from those affected strongly by beaming or by
obscuration of the nuclear emission.
Comparing bolometric luminosity to black hole mass for
234 AGN, we find little or no correlation.
Gaps in coverage of the $L_{bol}$--$M_{BH}$ plane are due
at least in part to high-mass, low-luminosity objects like the
BL Lac objects and radio galaxies for which we have no
good bolometric luminosity estimates.

For a given black hole mass, bolometric luminosities
range over as many as four orders of magnitude. 
The Eddington ratios span nearly as large a range,
2--3 orders of magnitude at most luminosities.
These are much larger than any uncertainties in the
estimates of either black hole mass or luminosity.
There are no strong trends of Eddington ratio with
luminosity, contrary to long-held preconceptions.
The absence of low Eddington ratios at high redshifts
(high luminosities) can be explained at least in part 
by selection effects in flux-limited surveys wherein 
highly sub-Eddington AGN disappear progressively at 
higher redshifts.

We also do not confirm previously reported trends of
radio luminosity with black hole mass, and while our 
results indicate a modest dependence of radio loudness on black hole mass,
selection effects may exaggerate or even produce this trend. 
On the whole, black hole mass seems to have remarkably little
to do with the appearance of active nuclei, either their luminosities
or radio power.

Of course, the present sample includes a randomly selected
mix of AGN, with black hole masses estimated in different
ways, by different people, from different data sets.
There may be real trends dependent on other variables 
not taken into account here (e.g., AGN type).
It is obviously of interest to apply the more robust black
hole mass estimation methods ---
reverberation mapping and stellar velocity dispersion ---
to a large sample of AGN, at as high a redshift as possible,
although these methods will probably not work for
the typical AGN at $z\sim 2$--3.
In practice, such a study would start with measurements of
stellar velocity dispersions at $0.05 \lesssim z \lesssim 0.4$,
which require 4- to 10-m class telescopes.

\acknowledgements
We thank Matthew O'Dowd for suggesting the fundamental plane
method of estimating black hole masses and for helpful
discussions. We thank Aaron Barth for his careful reading
of the manuscript, and Meredith Hughes for help with the research.
Support for proposals 5938, 5939, 6363, and 7893 was provided by NASA 
through grants from the Space Telescope Science 
Institute, which is operated by the Association of
Universities for Research in Astronomy, Inc., 
under NASA contract NAS 5-26555.

\clearpage

\clearpage

\clearpage

\clearpage

\clearpage

\clearpage
\begin{figure}
\epsscale{.7}
\plotone{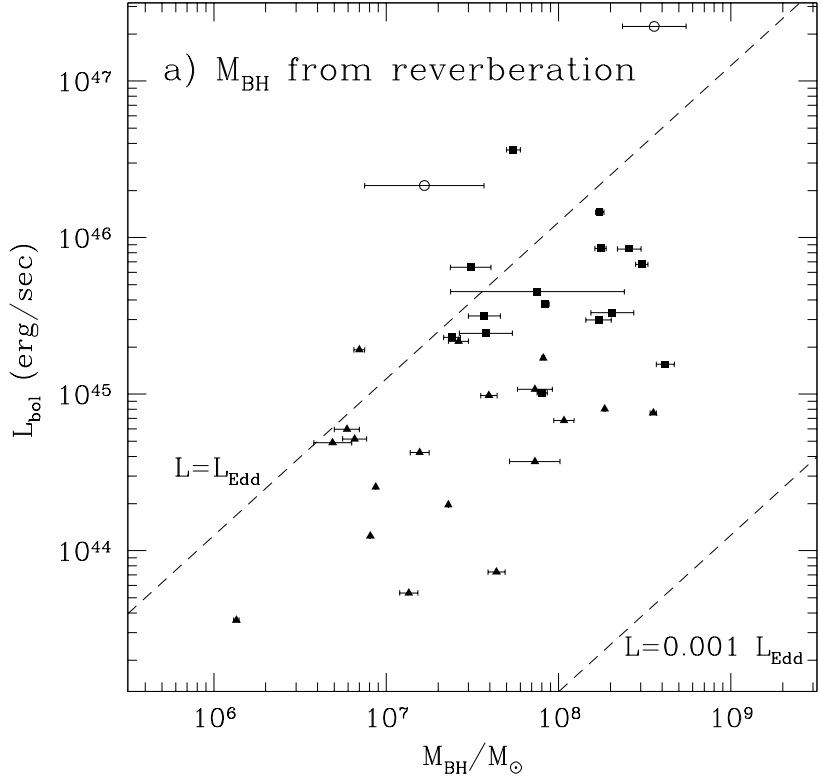}
\caption{
{\it a:}
Bolometric luminosity versus black hole mass for 36 
reverberation-mapped AGN.
The range in luminosity is roughly two orders of 
magnitude for a given black hole mass.
The mass plotted is the logarithmic mean from estimates with 
different velocity assumptions (measuring the FWHM from the 
rms spectrum or using the mean of FWHM measured from individual spectra),
with the error bar indicating the range.
{\it b:} 
The same mass--luminosity relation for Seyfert galaxies for which
black hole masses have been estimated from measured stellar 
velocity dispersions (Eq.~\ref{sig}). 
The bolometric luminosities of these Seyferts span 1-3 orders
of magnitude for a given black hole mass. 
The error bar indicates the uncertainty in
black hole mass due to the measurement error in $\sigma$.
{\it c:}
Mass--luminosity relation for 139 quasars whose black hole masses
have been estimated using line widths plus the optical luminosity 
to infer broad-line-region size
(McLure \& Dunlop 2001, Laor 2001, Gu et al. 2001).
A correlation is induced by the mass determination
if bolometric luminosity is linearly correlated with optical luminosity;
the correlation should follow 
$M_{BH} \propto L_{bol}^{0.7}$ ({\it thick line}).
Symbols are {\it open circles:} radio-loud quasars; {\it filled squares:} radio-quiet
quasars; {\it filled triangles:} Seyfert 1; {\it filled pentagons:}
Seyfert 2.}
\label{LBH1}
\end{figure}

\clearpage
\begin{figure}
\epsscale{1.0}
\plotone{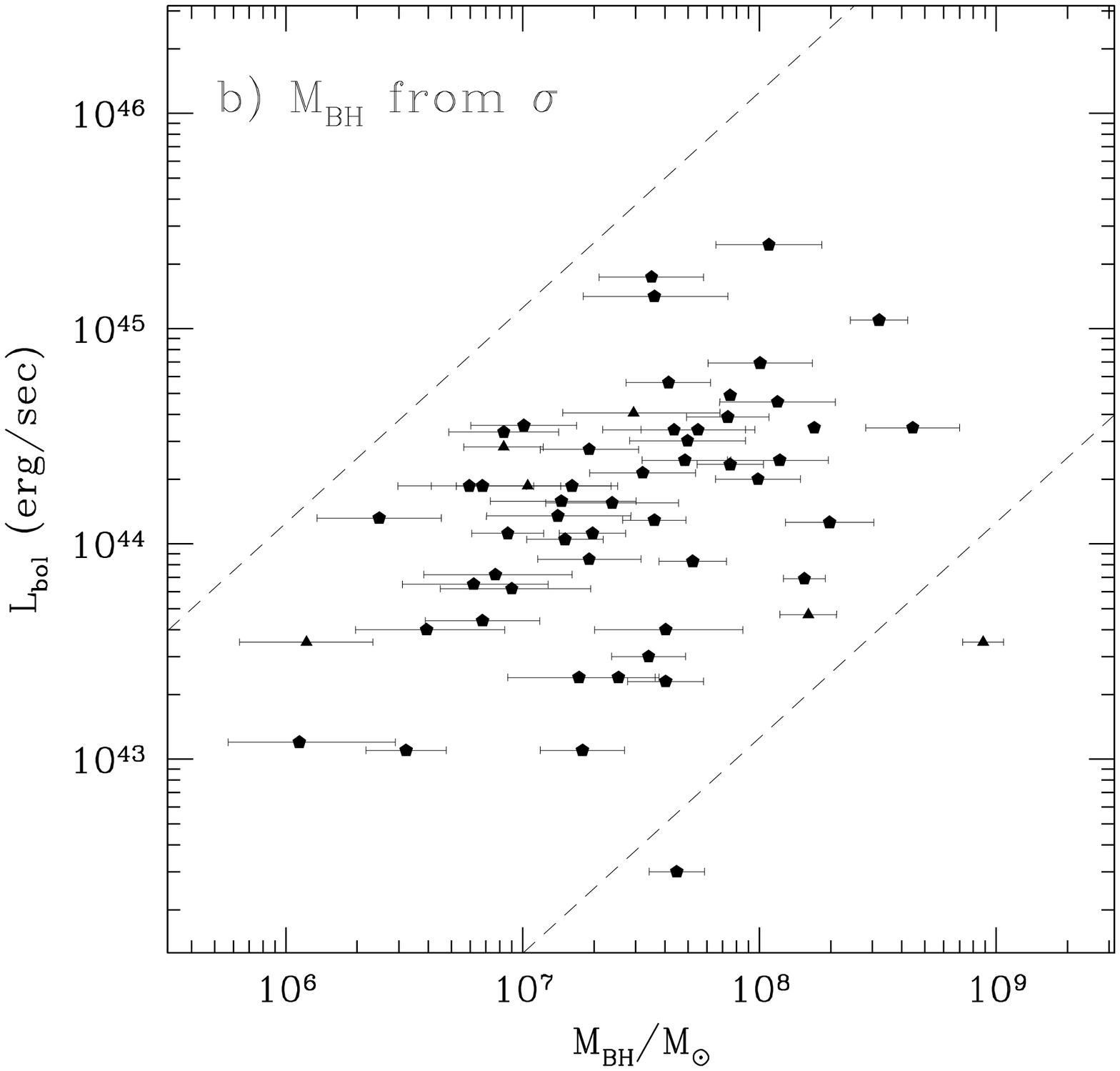}
\end{figure}

\clearpage
\begin{figure}
\plotone{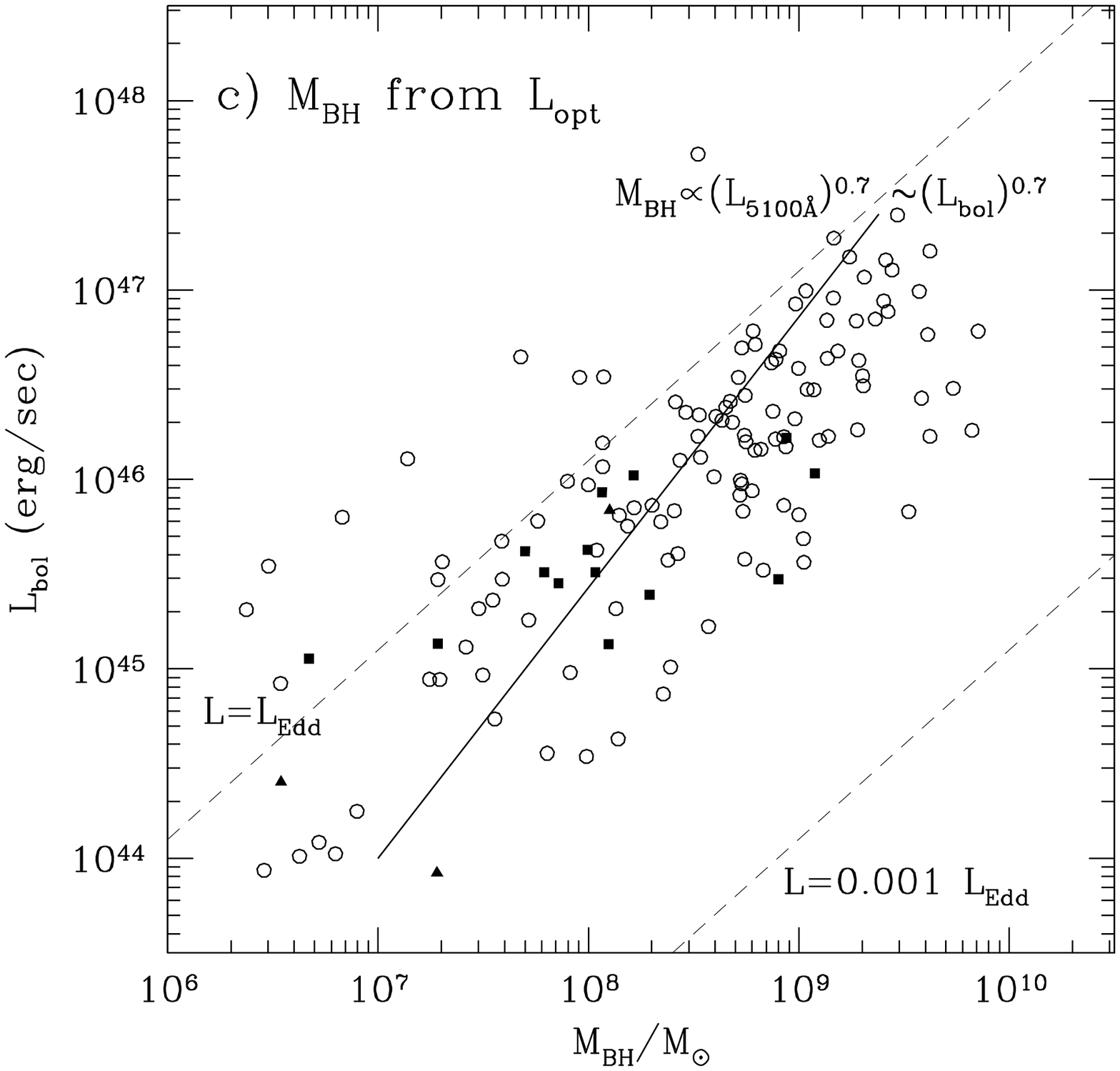}
\end{figure}

\clearpage
\begin{figure}
\plotone{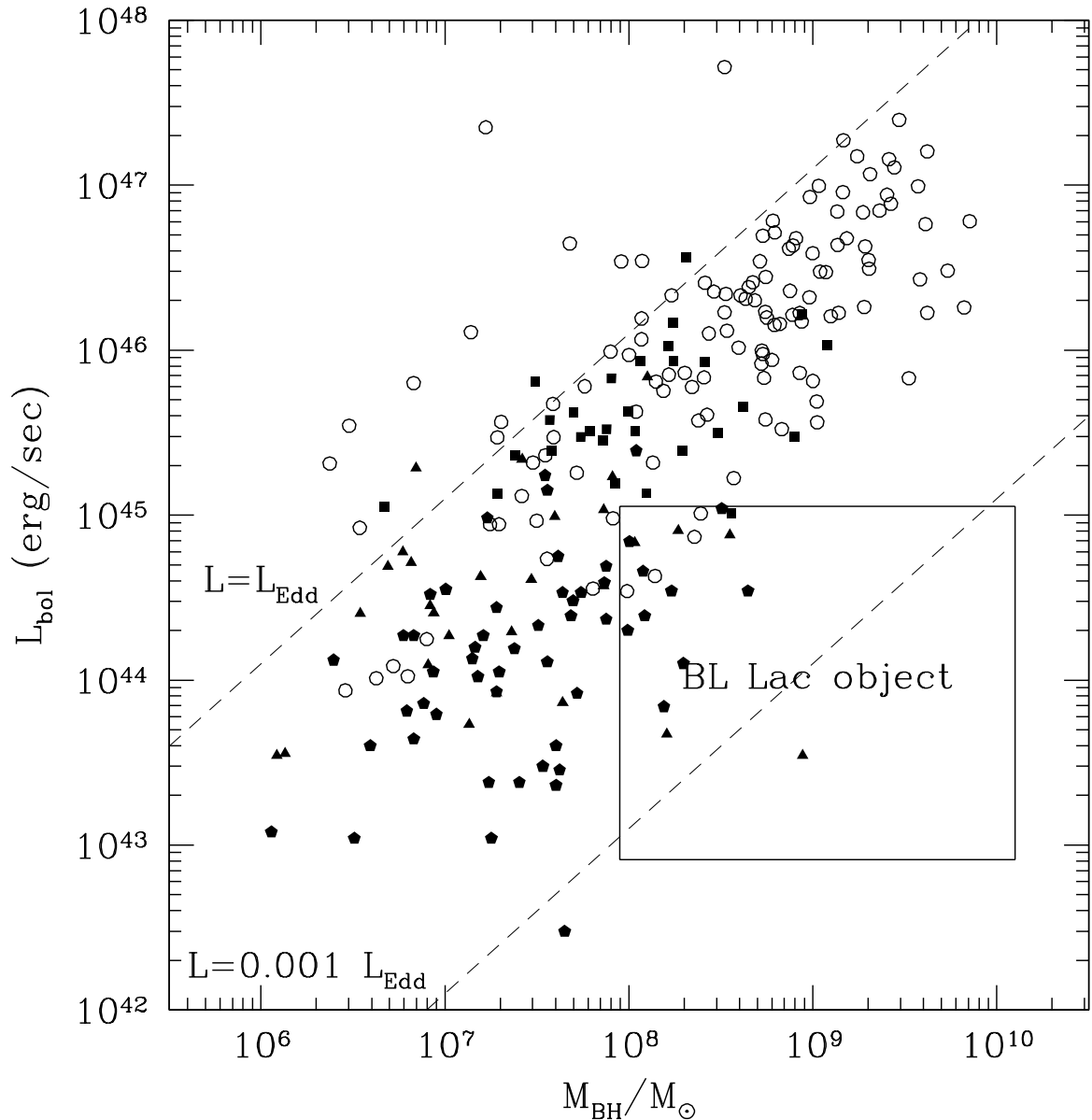}
\caption{Bolometric luminosity versus black hole mass for 234 AGN.
There is little if any correlation. 
For a given black hole mass, there is a large range of bolometric 
luminosities, spanning three or more orders of magnitude.
The Eddington limit defines an approximate upper limit 
to the luminosity,
but the absence of objects from the lower right of the diagram
(low luminosity, high mass AGN) is a selection effect. 
For example, this part of the diagram would be occupied by
BL Lac objects and low-luminosity radio galaxies. The inner box
indicates the approximate location of
BL Lac objects (see text). The symbols are the same as Figure 6.
\label{ML_ALL}
}
\end{figure}

\clearpage
\begin{figure}
\plotone{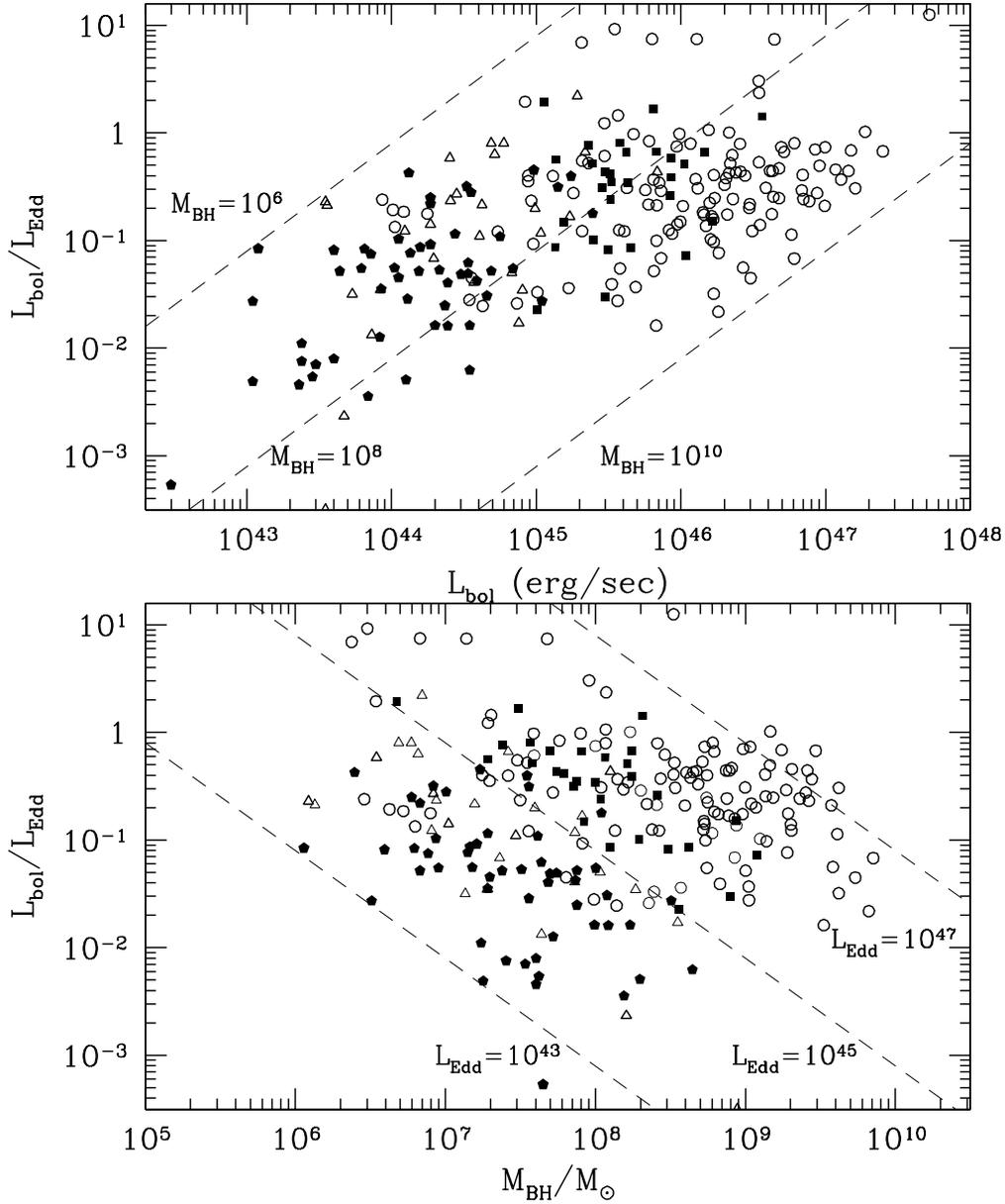}
\caption{
Eddington ratio versus bolometric luminosity ({\it top panel})
and versus black hole mass ({\it bottom panel}).
The range of Eddington ratios is roughly two orders of
magnitude over most of the observed luminosity or
black hole mass ranges.
The apparent deficit of high-luminosity objects
with low Eddington ratios (i.e., with black holes
in the range $10^{8} < M_{\rm BH} / M_\odot < 10^{10}$)
and of low-luminosity objects with high Eddington ratios,
as well as the absence of higher and lower luminosity 
AGN in the lower panel, 
are likely caused by selection effects
(see text). The symbols are the same as Figure 6.
\label{L_ER}
}
\end{figure}

\clearpage
\begin{figure}
\plotone{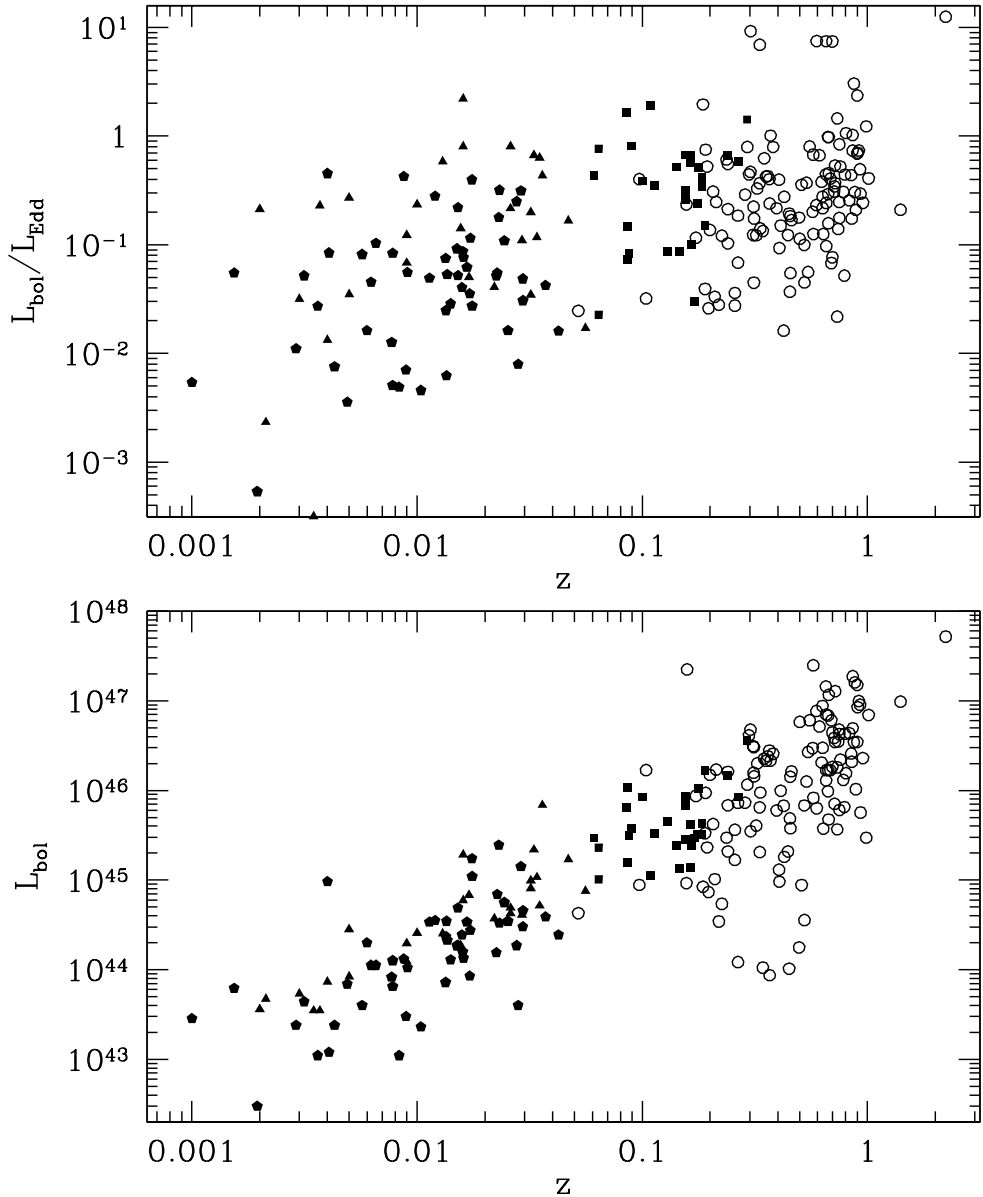}
\caption{
Eddington ratio ({\it top panel}) and bolometric luminosity 
({\it bottom panel}) versus redshift.
The Eddington ratio ranges from 0.001 to 1 at low redshifts, 
and from 0.01 to 10 at higher redshifts;  
although this represents a broad trend toward higher ratios 
at higher luminosities, the scatter is large and
selection effects are significant.
The bottom panel shows clearly selection effects that are
limiting the sample of AGN: the flux limit (lower envelope)
and the steepness of the luminosity function, which describes
how luminous objects more rare and thus are found only in
larger volumes, i.e., at higher redshifts (upper envelope).
These effects cause the broad distribution of Eddington ratios
in the top panel to be bounded, most notably in the lower
left. Even at that, the Eddington ratio has a broad range of
values at every redshift. The symbols are the same as Figure 6.
\label{Z_ER}
}
\end{figure}

\clearpage
\begin{figure}
\plotone{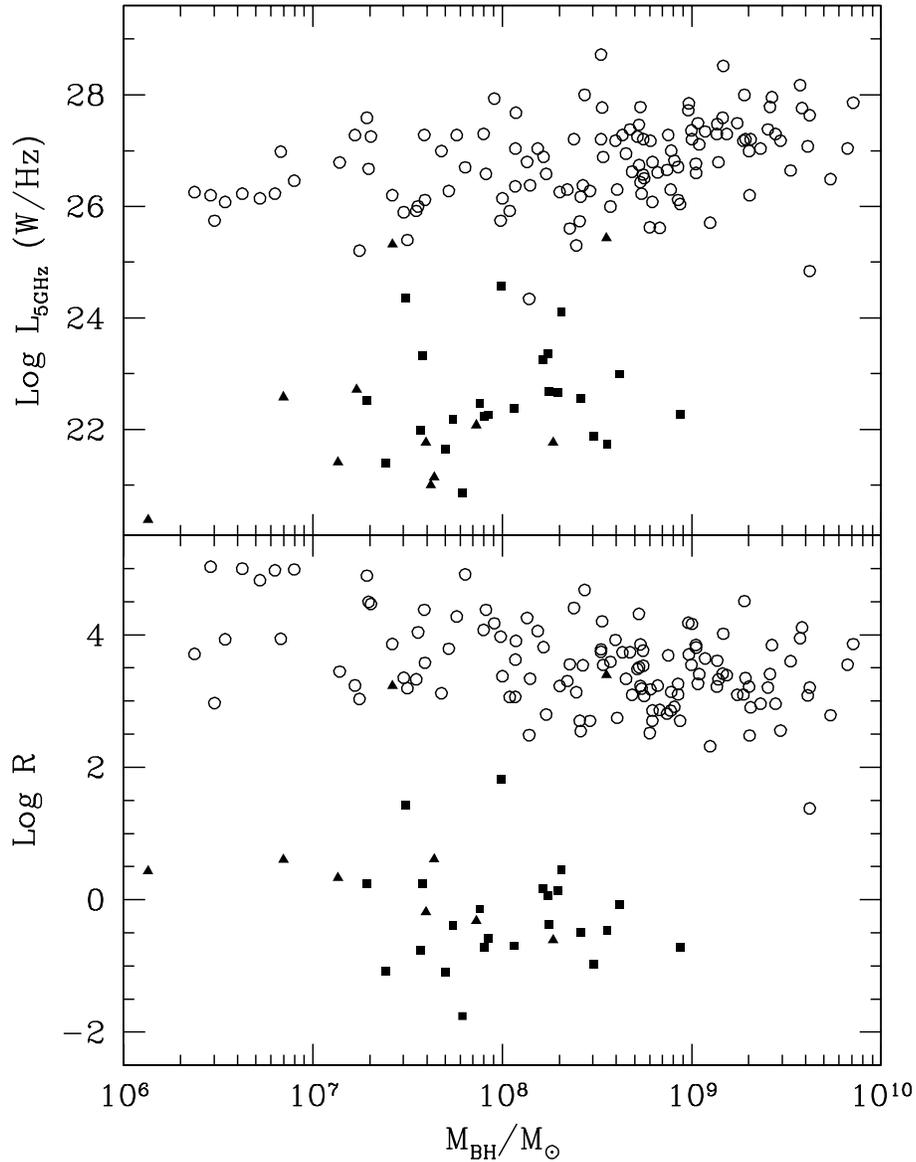}
\caption{Dependence of radio properties on black hole mass.
{\it Top panel:} Radio luminosity at 5~GHz versus black hole
mass for 157 AGN.
Both radio-quiet and radio-loud quasars span a large range in
black hole masses. The highest mass objects plotted do have
the highest radio luminosities but objects that would fall in
the lower right of the plot (BL Lac objects and radio galaxies)
have been excluded (due to the difficulty in accurately estimating
bolometric luminosity).
Note also that some of the highest radio power sources have some of
the lowest black hole masses. 
{\it Bottom panel:} 
Radio loudness ($f_{5GHz}/f_{opt}$) versus black hole mass for
the same 157 AGN.
There is little dependence of radio loudness on mass, apart from
an absence of the highest mass black holes in the radio-quiet
population; present data are not sufficient to determine whether
this absence is a real effect or due to sample selection and
observational bias. The symbols are the same as Figure 6.
\label{L5GHZ}
}
\end{figure}

\clearpage
\begin{deluxetable}{lcr}
\rotate
\tablewidth{0pt}
\tablecaption{Summary of Black Hole Mass Estimates}
\tablehead{
\colhead{Method} &
\colhead{Number} &
\colhead{References} }
\startdata
Spatially resolved kinematics & 2  & Greenhill et al. 1996, Miyoshi et al. 1995 \\
Reverberation mapping & 36  & Ho 1999, Kaspi et al. 2000, Onken \& Peterson 2002 \\
$L_{opt}$--$R_{BLR}$ relation& 139 & McLure \& Dunlop 2001, Laor 2001, Gu et al. 2001, Oshlack et al. 2002\\
$M_{BH}$--$\sigma$ relation & 33 &  Wu \& Han 2001, Barth et al. 2002, Falomo et al. 2002  \\
                            & 108  &  This work  \\
Fundamental plane & 59  &  This work \\ 
\enddata
\label{T_SUM}
\end{deluxetable}

\clearpage
\begin{deluxetable}{lcrrrrrr}
\tablewidth{0pt}
\tablecaption{Black Hole Masses from Spatially Resolved Kinematics}
\tablehead{
\colhead{Name} &
\colhead{z} &
\colhead{$L_{bol}$} &
\colhead{}&
\colhead{$M_{BH}$} &
\colhead{ref.} &
\colhead{Type} }
\startdata
NGC  1068 & 0.004& 44.98&I&  7.23  &1& SY2\\
NGC  4258 & 0.001& 43.45&I&  7.62  &2& SY2\\
\enddata
\label{T_kin}
\tablenotetext{a}{
Column (1) Name,
(2) redshift,
(3) log of the bolometric luminosity (ergs s$^{-1}$), 
(4) method for bolometric luminosity estimation
(I: flux integration; F: SED fitting),
(5) log of black hole mass in solar masses estimated from maser kinematics, 
(6) reference for black hole mass estimation, and
(7) AGN type.}
\tablerefs{
(1) Greenhill et al. (1996),
(2) Miyoshi et al. (1995).
}
\end{deluxetable}

\clearpage
\begin{deluxetable}{lcrrrrrr}
\tablewidth{0pt}
\tablecaption{Black Hole Masses from Reverberation Mapping}
\tablehead{
\colhead{Name} &
\colhead{z} &
\colhead{$L_{bol}$} &
\colhead{}&
\colhead{$M_{BH}$} &
\colhead{ref.} &
\colhead{Type} }
\startdata
3C   120  & 0.033& 45.34&I&  7.42  &1& SY1 \\
3C   390.3& 0.056& 44.88&I&  8.55  &1& SY1 \\
Akn  120  & 0.032& 44.91&I&  8.27  &1& SY1 \\
F    9    & 0.047& 45.23&F&  7.91  &1& SY1 \\
IC   4329A& 0.016& 44.78&I&  6.77  &1& SY1 \\
Mrk  79   & 0.022& 44.57&I&  7.86  &1& SY1 \\
Mrk  110  & 0.035& 44.71&F&  6.82  &1& SY1 \\
Mrk  335  & 0.026& 44.69&I&  6.69  &1& SY1 \\
Mrk  509  & 0.034& 45.03&I&  7.86  &1& SY1 \\
Mrk  590  & 0.026& 44.63&I&  7.20  &1& SY1 \\
Mrk  817  & 0.032& 44.99&I&  7.60  &1& SY1 \\
NGC  3227 & 0.004& 43.86&I&  7.64  &1& SY1 \\
NGC  3516 & 0.009& 44.29&I&  7.36  &3& SY1\\
NGC  3783 & 0.010& 44.41&I&  6.94  &2& SY1 \\
NGC  4051 & 0.002& 43.56&I&  6.13  &1& SY1 \\
NGC  4151 & 0.003& 43.73&I&  7.13  &1& SY1 \\
NGC  4593 & 0.009& 44.09&I&  6.91  &3& SY1\\
NGC  5548 & 0.017& 44.83&I&  8.03  &1& SY1 \\
NGC  7469 & 0.016& 45.28&I&  6.84  &1& SY1 \\
PG   0026+129& 0.142& 45.39&I&  7.58  &1& RQQ \\
PG   0052+251& 0.155& 45.93&F&  8.41  &1& RQQ \\
PG   0804+761& 0.100& 45.93&F&  8.24  &1& RQQ \\
PG   0844+349& 0.064& 45.36&F&  7.38  &1& RQQ \\
PG   0953+414& 0.239& 46.16&F&  8.24  &1& RQQ \\
PG   1211+143& 0.085& 45.81&F&  7.49  &1& RQQ \\
PG   1229+204& 0.064& 45.01&I&  8.56  &1& RQQ \\
PG   1307+085& 0.155& 45.83&F&  7.90  &1& RQQ \\
PG   1351+640& 0.087& 45.50&I&  8.48  &1& RQQ \\
PG   1411+442& 0.089& 45.58&F&  7.57  &1& RQQ \\
PG   1426+015& 0.086& 45.19&I&  7.92  &1& RQQ \\
PG   1613+658& 0.129& 45.66&I&  8.62  &1& RQQ \\
PG   1617+175& 0.114& 45.52&F&  7.88  &1& RQQ \\
PG   1700+518& 0.292& 46.56&F&  8.31  &1& RQQ \\
PG   2130+099& 0.061& 45.47&I&  7.74  &1& RQQ \\
PG   1226+023& 0.158& 47.35&I&  7.22  &1& RLQ \\
PG   1704+608& 0.371& 46.33&I&  8.23  &1& RLQ \\
\enddata
\label{T_rev}
\tablenotetext{a}{
Column (1) Name,
(2) redshift,
(3) log of the bolometric luminosity (ergs s$^{-1}$), 
(4) method for bolometric luminosity estimation
(I: flux integration; F: SED fitting),
(5) black hole mass estimate from reverberation mapping 
(for Kaspi et al. (2000) sample, where
black hole mass is log mean of rms FWHM and mean FWHM mass, in solar masses),
(6) reference for black hole mass estimation, and
(7) AGN type.}
\tablerefs{
(1) Kaspi et al. (2000),
(2) Onken \& Peterson (2002),
(3) Ho (1999).
}
\end{deluxetable}

\clearpage
\begin{deluxetable}{lcrrrrr}
\tablewidth{0pt}
\tabletypesize{\small}
\tablecaption{Black Hole Masses from Optical Luminosity}
\tablehead{
\colhead{Name} &
\colhead{z} &
\colhead{$L_{bol}$} &
\colhead{} &
\colhead{$M_{BH}$} &
\colhead{ref.} &
\colhead{Type} }
\startdata
Mrk 841 & 0.036& 45.84&I&  8.10  &1& SY1 \\
NGC 4253& 0.013& 44.40&I&  6.54  &1& SY1 \\
NGC 6814& 0.005& 43.92&I&  7.28  &1& SY1 \\
0054+144& 0.171& 45.47&F&  8.90  &2& RQQ \\
0157+001& 0.164& 45.62&F&  7.70  &2& RQQ \\
0204+292& 0.109& 45.05&F&  6.67  &2& RQQ \\
0205+024& 0.155& 45.45&F&  7.86  &2& RQQ \\
0244+194& 0.176& 45.51&F&  8.03  &2& RQQ \\
0923+201& 0.190& 46.22&F&  8.94  &2& RQQ \\
1012+008& 0.185& 45.51&F&  7.79  &2& RQQ \\
1029-140& 0.086& 46.03&F&  9.08  &2& RQQ \\
1116+215& 0.177& 46.02&F&  8.21  &2& RQQ \\
1202+281& 0.165& 45.39&F&  8.29  &2& RQQ \\
1309+355& 0.184& 45.63&F&  8.00  &2& RQQ \\
1402+261& 0.164& 45.13&F&  7.29  &2& RQQ \\
1444+407& 0.267& 45.93&F&  8.06  &2& RQQ \\
1635+119& 0.146& 45.13&F&  8.10  &2& RQQ \\
0022-297& 0.406& 44.98&F&  7.91  &3& RLQ \\
0024+348& 0.333& 45.31&F&  6.37  &3& RLQ \\
0056-001& 0.717& 46.54&F&  8.71  &3& RLQ \\
0110+495& 0.395& 45.78&F&  8.34  &3& RLQ \\
0114+074& 0.343& 44.02&F&  6.80  &4& RLQ \\
0119+041& 0.637& 45.57&F&  8.38  &3& RLQ \\
0133+207& 0.425& 45.83&F&  9.52  &3& RLQ \\
0133+476& 0.859& 46.69&F&  8.73  &3& RLQ \\
0134+329& 0.367& 46.44&F&  8.74  &3& RLQ \\
0135-247& 0.831& 46.64&F&  9.13  &3& RLQ \\
0137+012& 0.258& 45.22&F&  8.57  &2& RLQ \\
0153-410& 0.226& 44.74&F&  7.56  &4& RLQ \\
0159-117& 0.669& 46.84&F&  9.27  &3& RLQ \\
0210+860& 0.186& 44.92&F&  6.54  &3& RLQ \\
0221+067& 0.510& 44.94&F&  7.29  &4& RLQ \\
0237-233& 2.224& 47.72&F&  8.52  &3& RLQ \\
0327-241& 0.888& 46.01&F&  8.60  &4& RLQ \\
0336-019& 0.852& 46.32&F&  8.98  &3& RLQ \\
0403-132& 0.571& 46.47&F&  9.07  &3& RLQ \\
0405-123& 0.574& 47.40&F&  9.47  &3& RLQ \\
0420-014& 0.915& 47.00&F&  9.03  &3& RLQ \\
0437+785& 0.454& 46.15&F&  8.79  &3& RLQ \\
0444+634& 0.781& 46.12&F&  8.53  &3& RLQ \\
0454-810& 0.444& 45.32&F&  8.13  &3& RLQ \\
0454+066& 0.405& 45.12&F&  7.42  &4& RLQ \\
0502+049& 0.954& 46.36&F&  8.88  &4& RLQ \\
0514-459& 0.194& 45.36&F&  7.55  &3& RLQ \\
0518+165& 0.759& 46.34&F&  8.53  &3& RLQ \\
0538+498& 0.545& 46.43&F&  9.58  &3& RLQ \\
0602-319& 0.452& 45.69&F&  9.02  &3& RLQ \\
0607-157& 0.324& 46.30&F&  8.68  &3& RLQ \\
0637-752& 0.654& 47.16&F&  9.41  &3& RLQ \\
0646+600& 0.455& 45.58&F&  8.74  &3& RLQ \\
0723+679& 0.846& 46.41&F&  8.67  &3& RLQ \\
0736+017& 0.191& 45.97&F&  8.00  &2& RLQ \\
0738+313& 0.631& 46.94&F&  9.40  &3& RLQ \\
0809+483& 0.871& 46.54&F&  7.96  &3& RLQ \\
0838+133& 0.684& 46.23&F&  8.52  &3& RLQ \\
0906+430& 0.668& 45.99&F&  7.90  &3& RLQ \\
0912+029& 0.427& 45.26&F&  7.72  &4& RLQ \\
0921-213& 0.052& 44.63&F&  8.14  &4& RLQ \\
0923+392& 0.698& 46.26&F&  9.28  &3& RLQ \\
0925-203& 0.348& 46.35&F&  8.46  &4& RLQ \\
0953+254& 0.712& 46.59&F&  9.00  &3& RLQ \\
0954+556& 0.901& 46.54&F&  8.07  &3& RLQ \\
1004+130& 0.240& 46.21&F&  9.10  &2& RLQ \\
1007+417& 0.612& 46.71&F&  8.79  &3& RLQ \\
1016-311& 0.794& 46.63&F&  8.89  &4& RLQ \\
1020-103& 0.197& 44.87&F&  8.36  &2& RLQ \\
1034-293& 0.312& 46.20&F&  8.75  &3& RLQ \\
1036-154& 0.525& 44.55&F&  7.80  &4& RLQ \\
1045-188& 0.595& 45.80&F&  6.83  &3& RLQ \\
1100+772& 0.311& 46.49&F&  9.31  &3& RLQ \\
1101-325& 0.355& 46.33&F&  8.61  &4& RLQ \\
1106+023& 0.157& 44.97&F&  7.50  &4& RLQ \\
1107-187& 0.497& 44.25&F&  6.90  &4& RLQ \\
1111+408& 0.734& 46.26&F&  9.82  &3& RLQ \\
1128-047& 0.266& 44.08&F&  6.72  &4& RLQ \\
1136-135& 0.554& 46.78&F&  8.78  &3& RLQ \\
1137+660& 0.656& 46.85&F&  9.36  &3& RLQ \\
1150+497& 0.334& 45.98&F&  8.73  &3& RLQ \\
1151-348& 0.258& 45.56&F&  9.02  &3& RLQ \\
1200-051& 0.381& 46.41&F&  8.41  &4& RLQ \\
1202-262& 0.789& 45.81&F&  9.00  &3& RLQ \\
1217+023& 0.240& 45.83&F&  8.41  &2& RLQ \\
1237-101& 0.751& 46.63&F&  9.28  &4& RLQ \\
1244-255& 0.633& 46.48&F&  9.04  &3& RLQ \\
1250+568& 0.321& 45.61&F&  8.42  &3& RLQ \\
1253-055& 0.536& 46.10&F&  8.43  &3& RLQ \\
1254-333& 0.190& 45.52&F&  8.83  &4& RLQ \\
1302-102& 0.286& 45.86&F&  8.30  &2& RLQ \\
1352-104& 0.332& 45.81&F&  8.15  &4& RLQ \\
1354+195& 0.720& 47.11&F&  9.44  &3& RLQ \\
1355-416& 0.313& 46.48&F&  9.73  &3& RLQ \\
1359-281& 0.803& 46.19&F&  8.07  &4& RLQ \\
1450-338& 0.368& 43.94&F&  6.46  &4& RLQ \\
1451-375& 0.314& 46.16&F&  8.82  &3& RLQ \\
1458+718& 0.905& 46.93&F&  8.98  &3& RLQ \\
1509+022& 0.219& 44.54&F&  7.99  &4& RLQ \\
1510-089& 0.361& 46.38&F&  8.65  &3& RLQ \\
1545+210& 0.266& 45.86&F&  8.93  &2& RLQ \\
1546+027& 0.412& 46.00&F&  8.72  &3& RLQ \\
1555-140& 0.097& 44.94&F&  7.25  &4& RLQ \\
1611+343& 1.401& 46.99&F&  9.57  &3& RLQ \\
1634+628& 0.988& 45.47&F&  7.28  &3& RLQ \\
1637+574& 0.750& 46.68&F&  9.18  &3& RLQ \\
1641+399& 0.594& 46.89&F&  9.42  &3& RLQ \\
1642+690& 0.751& 45.78&F&  7.76  &3& RLQ \\
1656+053& 0.879& 47.21&F&  9.62  &3& RLQ \\
1706+006& 0.449& 44.01&F&  6.63  &4& RLQ \\
1721+343& 0.206& 45.63&F&  8.04  &3& RLQ \\
1725+044& 0.293& 46.07&F&  8.07  &3& RLQ \\
1726+455& 0.714& 45.85&F&  8.22  &3& RLQ \\
1828+487& 0.691& 46.78&F&  9.85  &3& RLQ \\
1849+670& 0.657& 46.23&F&  9.14  &3& RLQ \\
1856+737& 0.460& 46.21&F&  8.89  &3& RLQ \\
1928+738& 0.302& 46.68&F&  8.91  &3& RLQ \\
1945+725& 0.303& 45.54&F&  6.48  &3& RLQ \\
1954-388& 0.626& 46.31&F&  8.63  &4& RLQ \\
2004-447& 0.240& 45.32&F&  7.48  &4& RLQ \\
2043+749& 0.104& 46.23&F&  9.62  &3& RLQ \\
2059+034& 1.012& 46.84&F&  9.13  &4& RLQ \\
2111+801& 0.524& 45.83&F&  8.73  &3& RLQ \\
2120+099& 0.932& 45.75&F&  8.19  &4& RLQ \\
2128-123& 0.501& 46.76&F&  9.61  &3& RLQ \\
2135-147& 0.200& 46.17&F&  8.94  &2& RLQ \\
2141+175& 0.213& 46.23&F&  8.74  &2& RLQ \\
2143-156& 0.698& 46.65&F&  7.68  &4& RLQ \\
2155-152& 0.672& 45.67&F&  7.59  &3& RLQ \\
2201+315& 0.298& 46.62&F&  8.87  &3& RLQ \\
2216-038& 0.901& 47.17&F&  9.24  &3& RLQ \\
2218+395& 0.655& 46.11&F&  7.14  &3& RLQ \\
2247+140& 0.237& 45.47&F&  7.59  &2& RLQ \\
2251+158& 0.859& 47.27&F&  9.17  &3& RLQ \\
2255-282& 0.926& 46.96&F&  9.16  &3& RLQ \\
2311+469& 0.741& 46.55&F&  9.30  &3& RLQ \\
2329-415& 0.671& 46.22&F&  8.93  &4& RLQ \\
2342+821& 0.735& 45.56&F&  7.31  &3& RLQ \\
2344+092& 0.673& 47.07&F&  9.31  &3& RLQ \\
2345-167& 0.576& 45.92&F&  8.72  &3& RLQ \\
2349-014& 0.173& 45.94&F&  8.78  &2& RLQ \\
2355-082& 0.210& 45.01&F&  8.39  &2& RLQ \\
\enddata
\label{T_opt}
\tablenotetext{a}{
Column (1) Name, 
(2) redshift,
(3) log of the bolometric luminosity (ergs s$^{-1}$), 
(4) method for bolometric luminosity estimation 
(I: flux integration; F: SED fitting),
(5) log of the black hole mass in solar masses,
estimated using $L_{opt}-M_{BH}$ relation (Eq.~\ref{Lopt}),
(6) reference for optical luminosity,
(7) AGN type.}
\tablerefs{
(1) Laor 2001,
(2) McLure \& Dunlop 2001,
(3) Gu et al. 2001,
(4) Oshlack et al. 2002.}
\end{deluxetable}

\clearpage
\begin{deluxetable}{lcrrrrrr}
\tablewidth{0pt}
\tablecaption{Black Hole Masses from Observed Stellar Velocity Dispersions}
\tablehead{
\colhead{Name} &
\colhead{z} &
\colhead{$\sigma$} &
\colhead{ref.} &
\colhead{$M_{BH}$} &
\colhead{$L_{bol}$}&
\colhead{} &
\colhead{Type} }
\startdata
NGC 1566   &   0.005&    100.&N&    6.92&   44.45&I&SY1\\
NGC 2841   &   0.002&    209.&N&    8.21&   43.67&I&SY1\\
NGC 3982   &   0.004&     62.&N&    6.09&   43.54&I&SY1\\
NGC 3998   &   0.003&    319.&N&    8.95&   43.54&I&SY1\\
Mrk 10     &   0.029&    137.&N&    7.47&   44.61&I&SY1\\
UGC 3223   &   0.016&    106.&N&    7.02&   44.27&I&SY1\\
NGC 513    &   0.002&    152.&N&    7.65&   42.52&I&SY2\\
NGC 788    &   0.014&    140.&N&    7.51&   44.33&I&SY2\\
NGC 1052   &   0.005&    207.&N&    8.19&   43.84&I&SY2\\
NGC 1275   &   0.018&    248.&N&    8.51&   45.04&I&SY2\\
NGC 1320   &   0.009&    116.&N&    7.18&   44.02&I&SY2\\
NGC 1358   &   0.013&    173.&N&    7.88&   44.37&I&SY2\\
NGC 1386   &   0.003&    120.&N&    7.24&   43.38&I&SY2\\
NGC 1667   &   0.015&    173.&N&    7.88&   44.69&I&SY2\\
NGC 2110   &   0.008&    220.&N&    8.30&   44.10&I&SY2\\
NGC 2273   &   0.006&    124.&N&    7.30&   44.05&I&SY2\\
NGC 2992   &   0.008&    158.&N&    7.72&   43.92&I&SY2\\
NGC 3185   &   0.004&     61.&N&    6.06&   43.08&I&SY2\\
NGC 3362   &   0.028&     92.&N&    6.77&   44.27&I&SY2\\
NGC 3786   &   0.009&    142.&N&    7.53&   43.47&I&SY2\\
NGC 4117   &   0.003&     95.&N&    6.83&   43.64&F&SY2\\
NGC 4339   &   0.004&    132.&N&    7.40&   43.38&I&SY2\\
NGC 5194   &   0.002&    102.&N&    6.95&   43.79&I&SY2\\
NGC 5252   &   0.023&    190.&N&    8.04&   45.39&F&SY2\\
NGC 5273   &   0.004&     79.&N&    6.51&   43.03&I&SY2\\
NGC 5347   &   0.008&     93.&N&    6.79&   43.81&I&SY2\\
NGC 5427   &   0.009&     74.&N&    6.39&   44.12&I&SY2\\
NGC 5929   &   0.008&    121.&N&    7.25&   43.04&I&SY2\\
NGC 5953   &   0.007&    101.&N&    6.94&   44.05&I&SY2\\
NGC 6104   &   0.028&    148.&N&    7.60&   43.60&I&SY2\\
NGC 7213   &   0.006&    185.&N&    7.99&   44.30&I&SY2\\
NGC 7319   &   0.023&    130.&N&    7.38&   44.19&I&SY2\\
NGC 7603   &   0.030&    194.&N&    8.08&   44.66&I&SY2\\
NGC 7672   &   0.013&     98.&N&    6.88&   43.86&I&SY2\\
NGC 7682   &   0.017&    123.&N&    7.28&   43.93&I&SY2\\
NGC 7743   &   0.006&     83.&N&    6.59&   43.60&I&SY2\\
Mrk 1      &   0.016&    115.&N&    7.16&   44.20&I&SY2\\
Mrk 3      &   0.014&    269.&N&    8.65&   44.54&I&SY2\\
Mrk 78     &   0.037&    172.&N&    7.87&   44.59&I&SY2\\
Mrk 270    &   0.010&    148.&N&    7.60&   43.37&I&SY2\\
Mrk 348    &   0.015&    118.&N&    7.21&   44.27&I&SY2\\
Mrk 533    &   0.029&    144.&N&    7.56&   45.15&I&SY2\\
Mrk 573    &   0.017&    123.&N&    7.28&   44.44&I&SY2\\
Mrk 622    &   0.023&    100.&N&    6.92&   44.52&I&SY2\\
Mrk 686    &   0.014&    144.&N&    7.56&   44.11&I&SY2\\
Mrk 917    &   0.024&    149.&N&    7.62&   44.75&I&SY2\\
Mrk 1018   &   0.042&    195.&N&    8.09&   44.39&I&SY2\\
Mrk 1040   &   0.017&    151.&N&    7.64&   44.53&I&SY2\\
Mrk 1066   &   0.012&    105.&N&    7.01&   44.55&I&SY2\\
Mrk 1157   &   0.015&     95.&N&    6.83&   44.27&I&SY2\\
Akn 79  &   0.018&    143.&N&    7.54&   45.24&F&SY2\\
Akn 347 &   0.023&    186.&N&    8.00&   44.84&F&SY2\\
IC 5063 &   0.011&    160.&N&    7.74&   44.53&I&SY2\\
II ZW55 &   0.025&    212.&N&    8.23&   44.54&F&SY2\\
F 341   &   0.016&    114.&N&    7.15&   44.13&I&SY2\\
UGC 3995 &   0.016&    155.&N&    7.69&   44.39&I&SY2\\
UGC 6100 &   0.029&    156.&N&    7.70&   44.48&I&SY2\\
1ES 1959+65 &0.048&    195.&F&    8.09&-&&BLL\\
Mrk 180 &   0.045&    209.&Ba&    8.21&-&&BLL\\
Mrk 421 &   0.031&    219.&Ba&    8.29&-&&BLL\\
Mrk 501 &   0.034&    372.&Ba&    9.21&-&&BLL\\
I Zw 187&   0.055&    171.&Ba&    7.86&-&&BLL\\
3C 371  &   0.051&    249.&Ba&    8.51&-&&BLL\\
1514-241&   0.049&    196.&Ba&    8.10&-&&BLL\\
0521-365&   0.055&    269.&Ba&    8.65&-&&BLL\\
0548-322&   0.069&    202.&Ba&    8.15&-&&BLL\\
0706+591&   0.125&    216.&Ba&    8.26&-&&BLL\\
2201+044&   0.027&    197.&Ba&    8.10&-&&BLL\\
2344+514&   0.044&    294.&Ba&    8.80&-&&BLL\\
3C 29    &   0.045&    208.&B&    8.20&-&&RG\\
3C 31    &   0.017&    248.&B&    8.50&-&&RG\\
3C 33    &   0.059&    230.&B&    8.38&-&&RG\\
3C 40    &   0.018&    171.&B&    7.86&-&&RG\\
3C 62    &   0.148&    273.&B&    8.67&-&&RG\\
3C 76.1  &   0.032&    200.&B&    8.13&-&&RG\\
3C 78    &   0.029&    261.&B&    8.60&-&&RG\\
3C 84    &   0.017&    246.&B&    8.49&-&&RG\\
3C 88    &   0.030&    189.&B&    8.03&-&&RG\\
3C 89    &   0.139&    250.&B&    8.52&-&&RG\\
3C 98    &   0.031&    173.&B&    7.88&-&&RG\\
3C 120   &   0.033&    200.&B&    8.13&-&&RG\\
3C 192   &   0.060&    192.&B&    8.06&-&&RG\\
3C 196.1 &   0.198&    210.&B&    8.21&-&&RG\\
3C 223   &   0.137&    202.&B&    8.15&-&&RG\\
3C 293   &   0.045&    185.&B&    7.99&-&&RG\\
3C 305   &   0.041&    178.&B&    7.92&-&&RG\\
3C 338   &   0.030&    290.&B&    8.78&-&&RG\\
3C 388   &   0.091&    365.&B&    9.18&-&&RG\\
3C 444   &   0.153&    155.&B&    7.68&-&&RG\\
3C 449   &   0.017&    224.&B&    8.33&-&&RG\\
gin 116  &   0.033&    285.&B&    8.75&-&&RG\\
NGC 315  &   0.017&    311.&B&    8.90&-&&RG\\
NGC 507  &   0.017&    329.&B&    9.00&-&&RG\\
NGC 708  &   0.016&    241.&B&    8.46&-&&RG\\
NGC 741  &   0.018&    280.&B&    8.72&-&&RG\\
NGC 4839 &   0.023&    244.&B&    8.48&-&&RG\\
NGC 4869 &   0.023&    199.&B&    8.12&-&&RG\\
NGC 4874 &   0.024&    266.&B&    8.63&-&&RG\\
NGC 6086 &   0.032&    322.&B&    8.96&-&&RG\\
NGC 6137 &   0.031&    295.&B&    8.81&-&&RG\\
NGC 7626 &   0.025&    324.&B&    8.97&-&&RG\\
0039-095&   0.000&    280.&B&    8.72&-&&RG\\
0053-015&   0.038&    297.&B&    8.82&-&&RG\\
0053-016&   0.043&    249.&B&    8.51&-&&RG\\
0055-016&   0.045&    302.&B&    8.85&-&&RG\\
0110+152&   0.044&    196.&B&    8.09&-&&RG\\
0112-000&   0.045&    252.&B&    8.53&-&&RG\\
0112+084&   0.000&    365.&B&    9.18&-&&RG\\
0147+360&   0.018&    242.&B&    8.46&-&&RG\\
0131-360&   0.030&    251.&B&    8.53&-&&RG\\
0257-398&   0.066&    219.&B&    8.29&-&&RG\\
0306+237&   0.000&    249.&B&    8.51&-&&RG\\
0312-343&   0.067&    257.&B&    8.57&-&&RG\\
0325+024&   0.030&    219.&B&    8.29&-&&RG\\
0431-133&   0.033&    269.&B&    8.65&-&&RG\\
0431-134&   0.035&    222.&B&    8.31&-&&RG\\
0449-175&   0.031&    158.&B&    7.72&-&&RG\\
0546-329&   0.037&    389.&B&    9.29&-&&RG\\
0548-317&   0.034&    123.&B&    7.28&-&&RG\\
0634-206&   0.056&    195.&B&    8.09&-&&RG\\
0718-340&   0.029&    331.&B&    9.01&-&&RG\\
0915-118&   0.054&    275.&B&    8.69&-&&RG\\
0940-304&   0.038&    389.&B&    9.29&-&&RG\\
1043-290&   0.060&    229.&B&    8.37&-&&RG\\
1107-372&   0.010&    295.&B&    8.81&-&&RG\\
1123-351&   0.032&    447.&B&    9.53&-&&RG\\
1258-321&   0.015&    263.&B&    8.61&-&&RG\\
1333-337&   0.013&    288.&B&    8.77&-&&RG\\
1400-337&   0.014&    309.&B&    8.89&-&&RG\\
1404-267&   0.022&    295.&B&    8.81&-&&RG\\
1510+076&   0.053&    336.&B&    9.03&-&&RG\\
1514+072&   0.035&    269.&B&    8.65&-&&RG\\
1520+087&   0.034&    220.&B&    8.29&-&&RG\\
1521-300&   0.020&    166.&B&    7.80&-&&RG\\
1602+178&   0.041&    213.&B&    8.24&-&&RG\\
1610+296&   0.032&    322.&B&    8.96&-&&RG\\
2236-176&   0.070&    245.&B&    8.49&-&&RG\\
2322+143&   0.045&    204.&B&    8.17&-&&RG\\
2322-122&   0.082&    224.&B&    8.33&-&&RG\\
2333-327&   0.052&    269.&B&    8.65&-&&RG\\
2335+267&   0.030&    345.&B&    9.08&-&&RG\\
\enddata
\label{T_sig}
\tablenotetext{a}{Column 
(1) name, 
(2) redshift, 
(3) stellar velocity dispersion (km s$^{-1}$), 
(4) reference for $\sigma$,
(5) black hole mass estimated using $M_{BH} \propto \sigma^{4.02}$ 
relation (Equation~\ref{sig}) in units of log $M_{\odot}$.
(6) Log of the bolometric luminosity (ergs s$^{-1}$). For BL Lac objects and
radio galaxies, bolometric luminosity is not estimated because of
uncertain effects of relativistic beaming and/or nuclear obscuration.
(7) Method for bolometric luminosity estimation (I: flux integration; F: SED fitting).
(8) AGN type: SY1: Seyfert 1; SY2: Seyfert 2; BLL: BL Lac object; RG: radio galaxy.}
\tablerefs{
N: Nelson (1995); F: Falomo et al. (2002); Ba: Barth et al. (2002); B: Bettoni et al. (2001)}
\end{deluxetable}

\clearpage
\begin{deluxetable}{lcrrrrrr}
\tablewidth{0pt}
\tablecaption{Black Hole Masses from Fundamental Plane-Derived Velocity Dispersions}
\tablehead{
\colhead{Name} &
\colhead{z} &
\colhead{$\mu_{1/2}$} &
\colhead{$r_{e}$(kpc)} &
\colhead{ref.}&
\colhead{$\sigma$(km/s)} &
\colhead{$M_{BH}$} &
\colhead{Type} }
\startdata
0122+090 &   0.339& 20.64&  4.13&1&  298.&  8.82&BLL\\
0145+138 &   0.124& 20.91&  3.43&1&  237.&  8.42&BLL\\
0158+001 &   0.229& 21.88&  5.87&1&  194.&  8.08&BLL\\
0229+200 &   0.139& 21.07&  6.97&1&  378.&  9.24&BLL\\
0257+342 &   0.247& 21.28&  5.68&1&  270.&  8.66&BLL\\
0317+183 &   0.190& 22.56&  8.82&1&  181.&  7.95&BLL\\
0331-362 &   0.308& 22.09& 11.54&1&  285.&  8.75&BLL\\
0347-121 &   0.188& 20.63&  3.37&1&  270.&  8.65&BLL\\
0350-371 &   0.165& 20.77&  4.16&1&  296.&  8.82&BLL\\
0414+009 &   0.287& 22.78& 16.78&1&  256.&  8.56&BLL\\
0419+194 &   0.512& 19.71&  1.91&1&  263.&  8.61&BLL\\
0506-039 &   0.304& 21.21&  5.91&1&  285.&  8.75&BLL\\
0525+713 &   0.249& 21.10&  6.46&1&  334.&  9.03&BLL\\
0607+710 &   0.267& 21.76&  8.19&1&  269.&  8.65&BLL\\
0737+744 &   0.315& 21.41&  7.92&1&  318.&  8.94&BLL\\
0922+749 &   0.638& 19.79&  4.40&1&  467.&  9.61&BLL\\
0927+500 &   0.188& 21.55&  5.39&1&  225.&  8.34&BLL\\
0958+210 &   0.344& 20.13&  3.25&1&  334.&  9.03&BLL\\
1104+384 &   0.031& 19.50&  2.25&1&  413.&  9.39&BLL\\
1133+161 &   0.460& 21.75&  7.09&1&  223.&  8.32&BLL\\
1136+704 &   0.045& 20.05&  2.50&1&  320.&  8.95&BLL\\
1207+394 &   0.615& 20.73&  6.14&1&  348.&  9.10&BLL\\
1212+078 &   0.136& 21.35&  7.17&1&  327.&  8.99&BLL\\
1215+303 &   0.130& 23.31& 16.98&1&  199.&  8.12&BLL\\
1218+304 &   0.182& 21.64&  6.84&1&  259.&  8.58&BLL\\
1221+245 &   0.218& 21.39&  3.73&1&  182.&  7.97&BLL\\
1229+643 &   0.164& 20.42&  4.87&1&  417.&  9.41&BLL\\
1248-296 &   0.370& 20.57&  4.53&1&  331.&  9.01&BLL\\
1255+244 &   0.141& 21.36&  5.42&1&  259.&  8.58&BLL\\
1407+595 &   0.495& 21.01&  8.26&1&  391.&  9.30&BLL\\
1418+546 &   0.152& 21.51&  8.39&1&  334.&  9.03&BLL\\
1426+428 &   0.129& 20.62&  4.55&1&  354.&  9.13&BLL\\
1440+122 &   0.162& 22.21&  9.41&1&  238.&  8.44&BLL\\
1534+014 &   0.312& 21.47&  7.50&1&  294.&  8.80&BLL\\
1704+604 &   0.280& 20.30&  2.99&1&  289.&  8.77&BLL\\
1728+502 &   0.055& 21.08&  3.06&1&  200.&  8.13&BLL\\
1757+703 &   0.407& 20.51&  3.67&1&  285.&  8.75&BLL\\
1807+698 &   0.051& 18.60&  1.90&1&  618.& 10.10&BLL\\
1853+671 &   0.212& 21.37&  4.40&1&  211.&  8.23&BLL\\
2005-489 &   0.071& 21.30&  6.89&1&  335.&  9.03&BLL\\
2143+070 &   0.237& 21.68&  6.64&1&  241.&  8.46&BLL\\
2200+420 &   0.069& 21.80&  5.71&1&  212.&  8.23&BLL\\
2254+074 &   0.190& 22.48& 13.29&1&  264.&  8.62&BLL\\
2326+174 &   0.213& 21.13&  5.29&1&  284.&  8.74&BLL\\
2356-309 &   0.165& 21.08&  4.52&1&  262.&  8.60&BLL\\
0230-027 &   0.239& 21.80&  5.13&2&  182.&  7.97&RG \\
0307+169 &   0.256& 21.40&  6.27&2&  271.&  8.66&RG \\
0345+337 &   0.244& 23.30&  8.73&2&  112.&  7.12&RG \\
0917+459 &   0.174& 23.00& 14.60&2&  209.&  8.21&RG \\
0958+291 &   0.185& 22.00&  5.67&2&  178.&  7.93&RG \\
1215-033 &   0.184& 22.00&  5.67&2&  179.&  7.93&RG \\
1215+013 &   0.118& 21.00&  3.13&2&  209.&  8.20&RG \\
1330+022 &   0.215& 22.90& 10.47&2&  167.&  7.82&RG \\
1342-016 &   0.167& 22.90& 15.53&2&  234.&  8.41&RG \\
2141+279 &   0.215& 23.50& 16.53&2&  168.&  7.82&RG \\
0257+024 &   0.115& 21.70&  7.80&2&  285.&  8.75&RQQ\\
1549+203 &   0.250& 22.20&  3.33&2&  100.&  6.92&RQQ\\
2215-037 &   0.241& 21.40&  4.47&2&  208.&  8.20&RQQ\\
2344+184 &   0.138& 23.80& 11.67&2&  109.&  7.07&RQQ\\
\enddata
\label{T_fp}
\tablenotetext{a}{Column 
(1) name, 
(2) redshift, 
(3) surface brightness at $r_{e}$ in the R band, 
(4) effective radius scaled with $H_0$=75~km s$^{-1}$,
(5) reference for original $\mu_{e}$ and $r_{e}$ 
(1=Urry et al. 2000, 2=Dunlop et al. 2002),
(6) stellar velocity dispersion esimated using  $\mu_{e}$ and $r_{e}$
(Eq. 5).
(7) log of black hole mass in solar masses estimated
from Eq.~\ref{sig}, with $\sigma$ derived from
$\mu_e$, $r_e$, and the fundamental plane relation,
(8) AGN type (BLL=BL Lac object, RG=radio galaxy, RQQ=radio-quiet quasar).
}
\end{deluxetable}


\clearpage

\begin{thebibliography}{}
\bibitem[Barth]{barth}Barth, A., Ho, L., \& Sargent, W. L. W. 2002, ApJ, submitted
\bibitem[Bettoni et al.]{bettoni01}Bettoni, D., Falomo, R., Fasano, G., Govoni, F., Salvo, M., \& Scarpa, R. 2001, A\&A, 380, 471
\bibitem[Blanford \& McKee]{blandford}Blandford, R. D., \& McKee, C. F. 1982, ApJ, 255, 419
\bibitem[Cardelli]{cardelli}Cardelli, J. A., Clayton, G. C., \& Mathis, J. 1989, ApJ, 345, 245
\bibitem[Cav]{cavaliere}Cavaliere, A., \& Padovani, P. 1989, ApJ, 340, L5
\bibitem[Dibai]{dibai81}Dibai, E. A. 1981, Soviet Astr., 24, 389
\bibitem[DiNel]{diN95}Di Nella, H., Garcia, A. M., Garnier, R., \& Paturel, G. 1995, A\&AS, 113, 151
\bibitem[Dondi]{dondi}Dondi, L., \& Ghisellini, G. 1995, MNRAS, 273, 583
\bibitem[Dunlop2002]{dunlop02}Dunlop, J. S., McLure, R. J., Kukula, M. J., Baum, S. A., O'Dea, C. P., \& Hughes, D. H. 2002, MNRAS, submitted (astro-ph/0108397)
\bibitem[Elvis et al.]{elvis94}Elvis, M., et al. 1994, ApJS, 95, 1
\bibitem[Falomo]{falomo}Falomo, R., Kotilainen, J. K., \& Treves, A. 2002, ApJ, 569, L35
\bibitem[Ferrarese]{ferrarese00}Ferrarese, L., \& Merritt, D. 2000, ApJ, 539, L9
\bibitem[Ferrarese et al.]{f01}Ferrarese, L., et al. 2001, ApJ, 555, L79
\bibitem[Fossati et al.]{fosa}Fossati, G., Maraschi, L., Celotti, A., Comastri, A. \& Ghisellini, G. 1998, MNRAS, 299, 433
\bibitem[Gebhardt et al.]{gebhardt00a}Gebhardt, K., et al. 2000a, ApJ, 539, L13
\bibitem[Gebhardt et al.]{gebhardt00}Gebhardt, K., et al. 2000b, ApJ, 543, L5
\bibitem[Greenhill et al.]{greenhill}Greenhill, L. J., et al. 1996, ApJ, 481, L23
\bibitem[Gu et al]{gu01}Gu, M., Cao, X., \& Jiang, D. R. 2001, MNRAS, 327, 1111
\bibitem[Harms et al.]{hamrs94}Harms, R. J. et al. 1994, ApJ, 435, L35
\bibitem[Ho 99]{ho99}Ho, L. C. 1999, in Observational Evidence for Black Holes in the Universe, ed. S. K. Chakrabarti (Dordrecht: Kluwer), 157 
\bibitem[Ho 02]{ho02}Ho, L. C. 2002, ApJ, 564, 120
\bibitem[Jorgensen et al.]{jorgensen96}Jorgensen, I., Franx, M., \& Kjargaard, P. 1996, MNRAS, 280, 167
\bibitem[Kaspi et al. 1996]{kaspi96}Kaspi, S., Smith, P. S., Maoz, D., Netzer, H., \& Jannuzi, B. T. 1996, ApJ, 471, L75 
\bibitem[Kaspi et al. 2000]{kaspi00}Kaspi, S., et al. 2000, ApJ, 533, 631
\bibitem[Koratkar1991]{koratkar91}Koratkar, A. P., \& C. M. Gaskell 1991, ApJ, 370, L61
\bibitem[Krolik 1991]{krolik91}Krolik, J. H., et al. 1991, ApJ, 371, 541
\bibitem[Krolik 2001]{krolik01}Krolik, J. H. 2001, ApJ, 551, 72 
\bibitem[Kormendy 2001]{kormendyr98}Kormendy, J., \& Gebhardt, K. 2001,
in The 20th Texas Symposium on Relativistic Astrophysics, ed. H. Martel \& 
J. C. Whleer, in press (astro-ph/0105230)
\bibitem[Lacy 2001]{lacy01}Lacy, M., Laurent-Meuleisen, S. A., Ridgway, S. E., Becker, R. H., \& White, R. L. 2001, ApJ, 551, L17 
\bibitem[Laor 2000]{laor00}Laor, A. 2000, ApJ, 543, L111
\bibitem[Laor 2001]{laor01}Laor, A. 2001, ApJ, 553, 677
\bibitem[Lynden-Bell]{lin69}Lynden-Bell, D. 1969, Nature, 223, 690
\bibitem[McLeod]{mcleod}McLeod, K. K., Rieke, G. H., \& Storrie-Lombardi, L. J.
1999 ApJ, 511, L67
\bibitem[McLure et al. 1999]{metal99}McLure, R. J., Kukula, M. J., Dunlop, J. S., Baum, S. A., \& O'Dea, C. P. 1999, MNRAS, 308, 377
\bibitem[McLure \& Dunlop 2001]{md01}McLure, R. J., \& Dunlop, J. S. 2001, MNRAS, 327, 199
\bibitem[McLure \& Dunlop 2001]{mj02}McLure, R. J., \& Jarvis, M. J. 2002, MNRAS, submitted (astroph 0204473)
\bibitem[Mas-Hesse et al.]{mas}Mas-Hesse, J. M., et al. 1995 A\&A, 298, 22
\bibitem[Miyoshi et al.]{miy}Miyoshi, M., et al. 1995, Nature, 373, 127
\bibitem[Nelson]{nel}Nelson, C. \& Whittle, M. 1995, ApJS, 99, 67
\bibitem[Netzer]{net}Netzer, H. 1990, in Active Galactic Nuclei: 1990 Sass-Fee Lectures, ed. T. J.-L. Courvoisier \& M. Mayor (Berlin: Springer), 137
\bibitem[Schmitt et al.]{schmitt}Schmitt, H., et al. 1997, AJ, 114, 592
\bibitem[O'Dowd et al.]{odowd}O'Dowd, M., C. M., Urry, \& R. Scarpa 2001, ApJ,
submitted
\bibitem[Oliva95]{oo95}Oliva, E., Origlia, L., Kotilainen, J.K., Moorwood, A.F.M. 1995, A\&A 301, 55
\bibitem[Oliva99]{oo99}Oliva, E., Origlia, L., Maiolino, R., Moorwood, A. F. M. 1999 A\&A 350, 9
\bibitem[Onken]{onken}Onken, C. A. \& Peterson, B. A. 2002, ApJ, acce2002, ApJ, in press
\bibitem[Oshlack]{oshlack}Oshlack, A., Webster, R., \& Whiting, M. 2002, PASP,
in press
\bibitem[Padovani]{padovani}Padovani, P., \& Rafanelli, P. 1988, A\&A 205, 53
\bibitem[Peterson88]{peterson88}Peterson, B. M. 1988, PASP, 100, 18
\bibitem[Peterson]{peterson}Peterson, B. M. 1993, PASP, 105, 247
\bibitem[Taylor]{taylor}Taylor, G. L., Dunlop, J. S., Hughes, D. H., \& Robson, E. I. 1996, MNRAS, 283, 930
\bibitem[Tremaine]{tremaine}Tremaine, S., et al. 2002, ApJ, in press (astro-ph/0203468)
\bibitem[Urry]{urry}Urry, C. M., et al. 2000, ApJ, 532, 816 
\bibitem[Vestergaard]{vest02}Vestergaard, M. 2002, ApJ, 571, 733
\bibitem[Wandel 1985]{wandel85}Wandel, A., \& Yahil, A. 1985, ApJ, 295, L1
\bibitem[Wandel 1999]{wandel99}Wandel, A., Peterson, B. M., \& Makkan, M. A.
1999, ApJ, 526, 579
\bibitem[Wu2001]{wu01}Wu, X.-B., \& Han, J. L. 2001, A\&A, 380, 31
\end{thebibliography}
\end{document}